\documentclass[journal,compsoc]{IEEEtran}
\usepackage{blindtext}
\usepackage{graphicx}

\usepackage{mypackages}

\begin{document}
%
\title{Performance Analysis of Indoor mmWave Networks with Ceiling-Mounted Access Points}
%
%
%

\author{Fadhil Firyaguna,
        Jacek Kibi{\l}da,
        Carlo Galiotto,
        and~Nicola~Marchetti 
\IEEEcompsocitemizethanks{\IEEEcompsocthanksitem The authors are with CONNECT Centre, Trinity College Dublin, Dublin, Ireland. E-mail: \{firyaguf, kibildj, galiotc, nicola.marchetti\}@tcd.ie.}
\thanks{Date of publication Feb 07, 2020; date of current version Feb 4, 2020.}%
\thanks{This publication has emanated from the research conducted within the scope of \textit{NEMO (Enabling Cellular Networks to Exploit Millimeter-wave Opportunities)} project financially supported by the Science Foundation Ireland (SFI) under Grant No. 14/US/I3110 and with partial support of the European Regional Development Fund under Grant No. 13/RC/2077. Earlier version of this paper was presented at the 2017 IEEE GLOBECOM, Singapore, December 7, 2017.}%
}

\IEEEtitleabstractindextext{%
\begin{abstract}
The objective of the Enhanced Mobile Broadband use case in 5G networks is to deliver high capacity access to densely populated areas, like city centres, transportation hubs or convention centres. Millimetre-wave communications are the go-to technology to realise that objective, yet due to weak outdoor-to-indoor penetration, outdoor deployments will not suffice and dedicated indoor deployments will be necessary.
In this article, we study dense deployments of millimetre-wave access points mounted on the ceiling, with directional antennas pointing downwards to illuminate selected spots on the ground. In this setup, the signal propagation is primarily limited by human body blockages.
Therefore, we develop a body blockage model and derive an expression for the probability of blockage. Using the developed expressions and our simulation framework, we assess the impact of densification and body blockage on the achievable performance.
We find that both coverage and area spectral efficiency curves exhibit non-trivial behaviour with respect to the access point density and that there is an optimal beamwidth-density configuration that only maximises either coverage or area spectral efficiency.
Such optimal configuration changes depending on the body blockage probability, leading to a necessity for network designers to carefully consider their intended application and scenario.

\end{abstract}

}

\maketitle

\IEEEdisplaynontitleabstractindextext

%
\IEEEpeerreviewmaketitle

\IEEEraisesectionheading{\section{Introduction}\label{sec:introduction}}
\IEEEPARstart{T}{o date}, the world's mobile industry has reached the 5 billion subscriber milestone, and in 2025, it is forecasted to reach 5.8 billion subscribers \cite{gsma2018mobileeconomy}.
ITU-R envisions that these mobile subscribers will be served by the Enhanced Mobile Broadband (eMBB) as part of the fifth generation (5G) of mobile networks \cite{shafi2017tutorial}. Some of whom will access the network through indoor hotspots in environments such as transportation hubs, sports arenas, and convention centres.
The eMBB will provide mobile services over multigigabit/s rate links, enabled by the high-density deployment of networks that utilise the spectrum available in \ac{mmWave} frequencies.
However, the application of mmWave frequencies to cellular networks requires a new set of assumptions to be taken into account in the system design.

\begin{figure}[t]
    \centering
    \includegraphics[width=.8\linewidth]{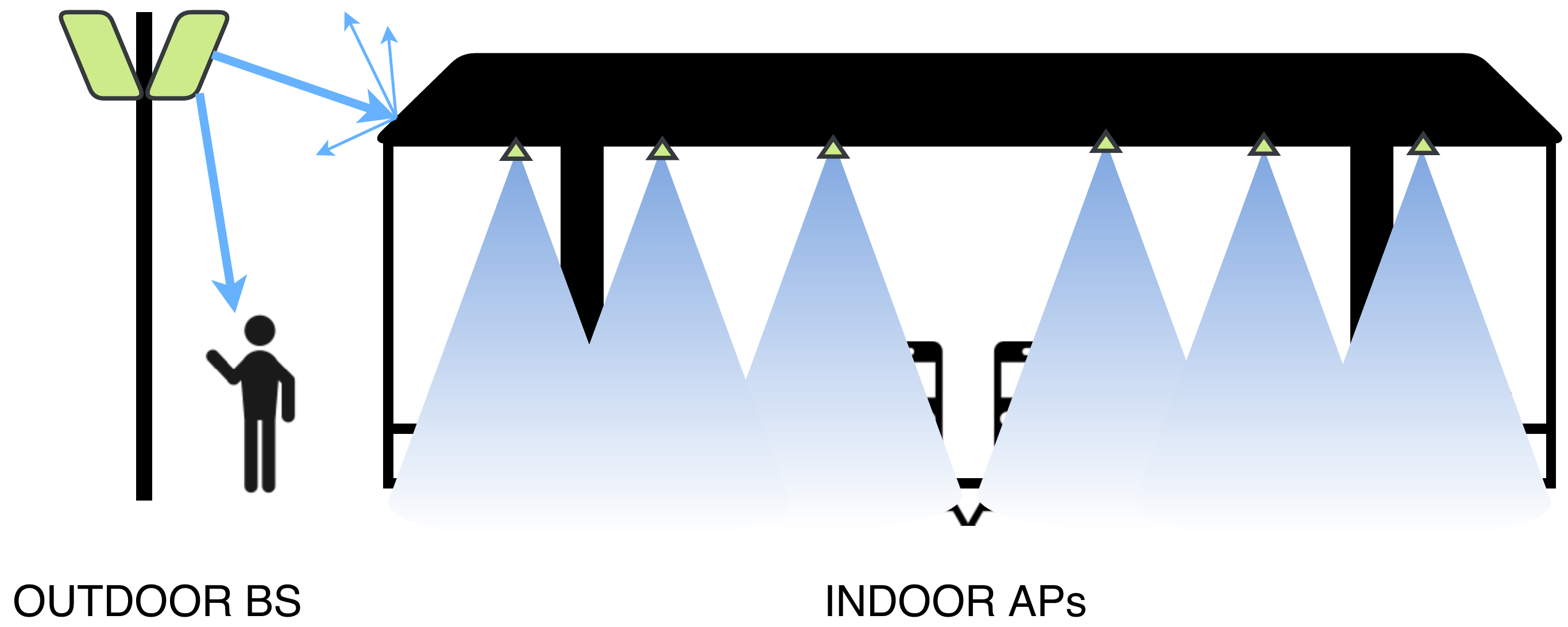}
    \caption{Independent tier of indoor \ac{mmWave} ceiling-mounted \ac{AP} deployment providing coverage in a transportation hub with directional transmission pointing downward. The \ac{mmWave} signal from outdoor base station (BS) does not propagate through building walls.}
    \label{fig:ceiling-mounted_deployment}
\end{figure}

The propagating mmWave signals suffer from high path attenuation and low penetration through materials such as building walls.
It means that mmWave signals transmitted from outdoor base stations will be confined to streets and other outdoor areas \cite{PiKhan_2011}.
This creates a situation where an independent tier of mmWave \acp{AP} should be deployed to ensure coverage to \acp{UE} in indoor areas, as we illustrate in Fig. \ref{fig:ceiling-mounted_deployment}.
Yet, coverage can still be deteriorated by human bodies, as they introduce as much as 40 dB of attenuation \cite{collonge2004influence,lu2012modeling,bai2014analysisSelfBlockage}, which may be enough to lose the connection between \ac{AP} and \ac{UE}.

In order to combat the detrimental effects of blockages, one would be tempted to increase the \ac{AP} density to maintain connectivity \cite{bai2014analysisSelfBlockage}. However, here comes another challenge --- inter-cell interference.
Although the increase in network density (densification) may improve the network capacity by allowing for an increase in spatial reuse \cite{hwang2013holistic}, if not implemented carefully, it may lead to excessive interference causing an increase in interference and a deterioration of the system's performance.
In state-of-the-art literature, the interference mitigation can be achieved by inter-cell interference coordination (ICIC) techniques, which demand high signalling overhead \cite{yoon2018distance-based}.
One simpler way to reduce the interference is to restrict the signal propagation through the use of directional antennas mounted on the ceiling.
Thus, the main-lobe of the directional antenna beam is pointed downwards to confine the signal's power to a limited space.

Ceiling installations of mmWave access points have been considered in the literature, e.g., \cite{yiu2009empirical,dong2012link,leong2004robust}, yet proposed analysis was restricted to single link performance. 
While, in state-of-the-art literature, we can find coverage and throughput analysis of such deployments operating in microwave frequencies, e.g., \cite{Ho2016downward}, the same analysis for deployments operating in mmWaves becomes more complicated due to the susceptibility to blockages.
To address this, in this work we consider a large indoor scenario where APs operating in mmWave frequencies (\ac{mmWave} \acp{AP}) are mounted on the ceiling to form a grid-like pattern, and fixed directional transmissions are set to illuminate selected spots on the ground. We consider a confined venue, where we assume that there is no outdoor-to-indoor interference. 

We consider a channel model where human bodies are the main source of signal blockage. The shadowing effects, as well as multipath propagation, are modelled as fading, whose parameters depend on the body blockage, as well as the path loss parameters.
We align the parameters of our channel model with those coming from the empirical measurements by Yoo et al. in \cite{yoo2017measurements}.

Our objective is to study the effects of different blockage scenarios on the optimal system design parameters (such as \ac{AP} density and antenna main-lobe beamwidth) and on the network performance. For this purpose:
\begin{itemize}
    \item We develop a model of body blockage that matches the geometry of the ceiling-mounted \ac{AP} deployment, and we derive an analytical expression for the probability of blockage.
    \item We analyse the impact of both AP density and body blockage on the performance through our simulation framework. The simulation source code we developed is available online (see Appendix~\ref{ap:simulationcodes}).
\end{itemize}
Based on an extensive simulation campaign:
\begin{enumerate}
    \item We find that coverage and \ac{ASE} achieve peak performance when there is a joint optimal beamwidth configuration of the ceiling-mounted AP antennas facing downwards and the directional UE antennas facing the AP, for a given AP density.
    \item We find that there exists a trade-off in beamwidth-density configuration, in which the same configuration optimises either coverage or \ac{ASE}, but not both.
    \item We find that the optimal beamwidth-density configuration for coverage depends on the body blockage probability, as the configuration needs to compensate for increased interference, shadowing or path loss.
\end{enumerate}

In what follows, we provide an overview of the related literature, description of our system model, and in-depth analysis of the numerical results obtained, and explaining the lessons learnt on the design of dense indoor mmWave networks.

\section{Related Work}
\label{sec:relatedworks}
Ceiling-mounted AP with fixed-beam antennas is an option for quick and low-cost deployments as recommended by ITU-R \cite{itu-r_m.2376-0}, facilitating AP densification. 
Dense deployments have been initially investigated in \cite{Ho2016downward} for microwave frequencies. 
In that work, Ho et al. have shown that the ceiling-mounted deployment with directional antennas was able to provide gains in throughput four times higher than the deployment with omnidirectional antennas.
Herein, we investigate the performance of such deployment considering the use of mmWave communication.
Hence, the blockage effects on the mmWave signal should be taken into account in the investigation, as the works in \cite{yiu2009empirical,dong2012link,leong2004robust} have shown that the link performance from a single ceiling-mounted mmWave transmitter can be degraded by human body blockage.

Dense mmWave networks have been widely studied for outdoor scenarios  \cite{andrews2017modeling,bai2014analysisBlockageUrban,bai2015coverage,bai2014analysisSelfBlockage} and much of the blockage modelling and their results are used for studying indoor scenarios as well.
In \cite{bai2015coverage}, Bai and Heath have shown that there is a finite optimal AP density for coverage, with fixed AP and UE beamwidth configurations, which is given by the transition from a noise-limited regime to an interference-limited regime when increasing AP density.
In \cite{bai2014analysisSelfBlockage}, Bai and Heath have proposed a cone blocking model to characterise the blockage by the user's body, and they have shown that self-body blocking effects can reduce the average throughput of dense outdoor mmWave networks by about 10\%.
Furthermore, coverage performance in dense indoor mmWave networks has been studied in \cite{venugopal2016device,niknam2018interference}. They have analysed the blockage probability of multiple bodies besides the user body, modelling human body blockages as circles randomly placed in a finite-area.
In \cite{venugopal2016device}, Venugopal et al. have shown that, in device-to-device communication, interference in dense networks is significantly affected by blockage.
This has also been shown by Niknam et al. in \cite{niknam2018interference}, where their results for a point process AP deployment have shown that increasing blockage density improves coverage as the number of blocked interferers increases.
However, the analyses above have not considered the height difference between the AP and UE in the blockage modelling.

When considering the AP height, the geometry of the blocking objects becomes significant for the blockage analysis.
As demonstrated by Gapeyenko et al. in \cite{gapeyenko2016analysis}, the received signal strength can be maximised when an optimal AP antenna height is deployed. This antenna height depends on the height and diameter distribution of the blockers.
Also, considering both blocker and AP antenna height, there is an area around the AP where the blocker does not obstruct the link as the LOS goes over the obstruction, as observed in \cite{dong2012link,jain2019impact}.
These works use their analytical framework to study the optimal AP height or AP density that minimise the blockage probability. In our analysis, considering AP and UE directivity and taking the three-dimensional geometry of AP deployment and blockages into account, we study the optimal density and antenna configurations for the ceiling-mounted setup that maximises coverage and spectral efficiency.

In our previous work \cite{firyaguna2017coverage}, we analysed only the self-body blockage effects without any other blockage source, and no fading nor receiver directionality were assumed. In this paper, we build on our previous work by extending our model. Precisely, we assume the receiver UE has a directional antenna which aligns to the transmitter AP, and we assume the presence of large and small-scale fading. Also, we consider multiple bodies that can potentially block the signal from an \ac{AP} to the \ac{UE}, and we provide an analytical model for the blockage probability. 
Then, we show the effect of the ceiling-mounted deployment on network densification by investigating the impact of both AP and UE main-lobe beamwidths configuration on the inter-cell interference under body blockage conditions.

\section{System Model}
\label{sec:systemmodel}
In this section, we introduce our model of a mmWave indoor network. The considered environment is an indoor confined area where there is no interference from outdoor signals.
The \acp{AP} are deployed on a hexagonal grid, and they are installed on the ceiling at some height above the UE level, with fixed directional antennas illuminating the floor below. 
We consider a \ac{UE} randomly located in the square-shaped indoor venue. 
The AP that provides the desired signal to the UE is referred to as the serving AP, while the other APs are referred to as interfering APs.
In our notation, we use the subscript $_\mathrm{A}$ to denote the parameters related to the \ac{AP}, the subscript $_\mathrm{U}$ to denote the parameters related to the UE, and the subscript $_\mathrm{B}$ to denote the body parameters.
We assume human bodies are the main source of blockages for the mmWave signals and change the link state between \ac{LOS} and \ac{NLOS}.
Thus, we assume that both path loss and channel gain have different characteristics in each state.
These characteristics are modelled according to empirical channel measurements described in \cite{yoo2017measurements}.

\subsection{Directivity Gain}
\begin{figure}
\centering    
    \begin{subfigure}{\linewidth}
        \centering
        \includegraphics[scale=.3]{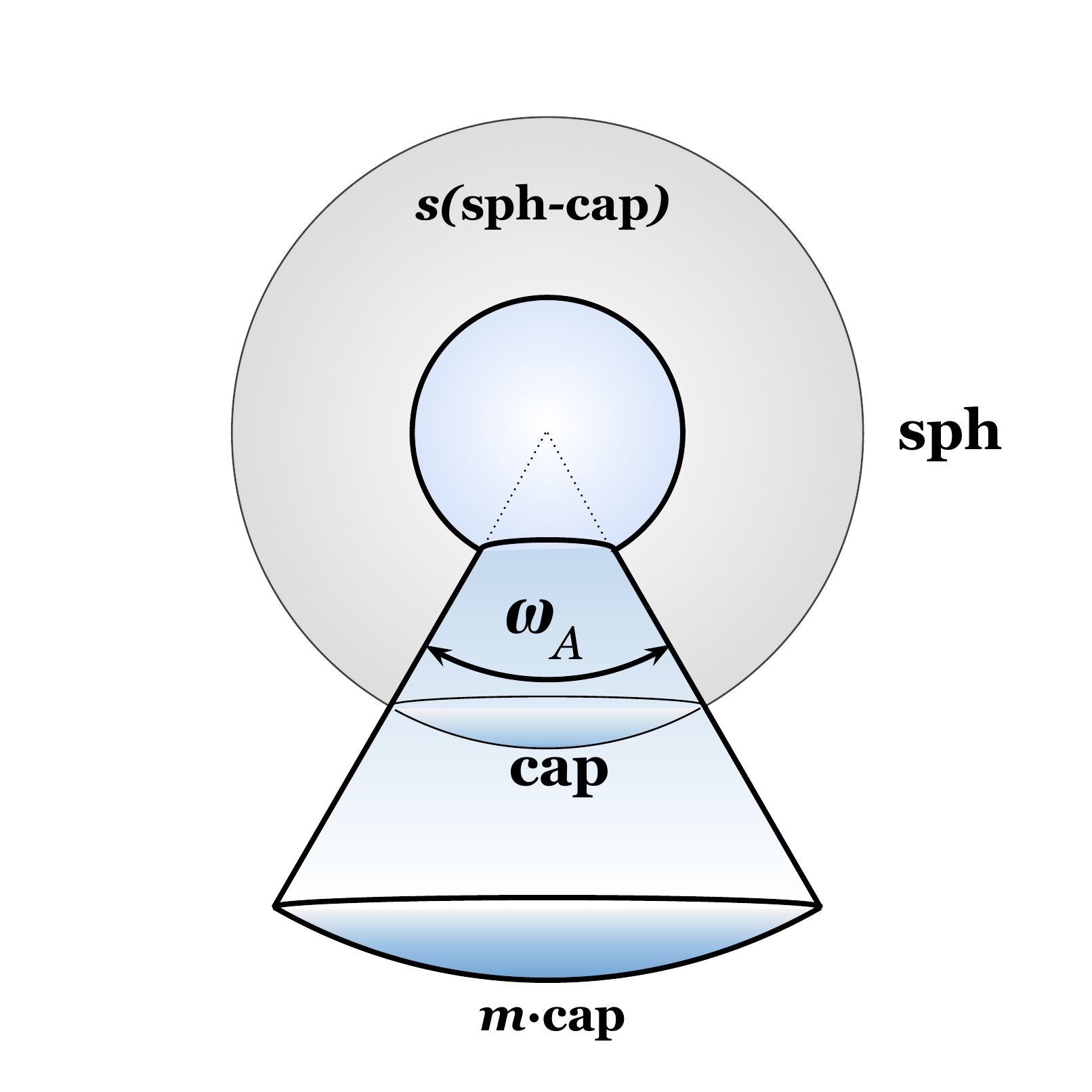}
        \caption{Cone-bulb model approximation of antenna directivity gain patterns. The cone represents the main-lobe and the bulb (inner sphere) represents the side-lobe. The cap area $\mathrm{cap}$ is scaled up by a factor of $m$, while the resulting bulb area $\mathrm{sph}-\mathrm{cap}$ is scaled down by a factor of $s$.}
        \label{fig:conebulbmodel}
    \end{subfigure}
    \hfill
    \begin{subfigure}{\linewidth}
        \centering
        \includegraphics[trim={0 0cm 0 3.5cm},clip,scale=.5]{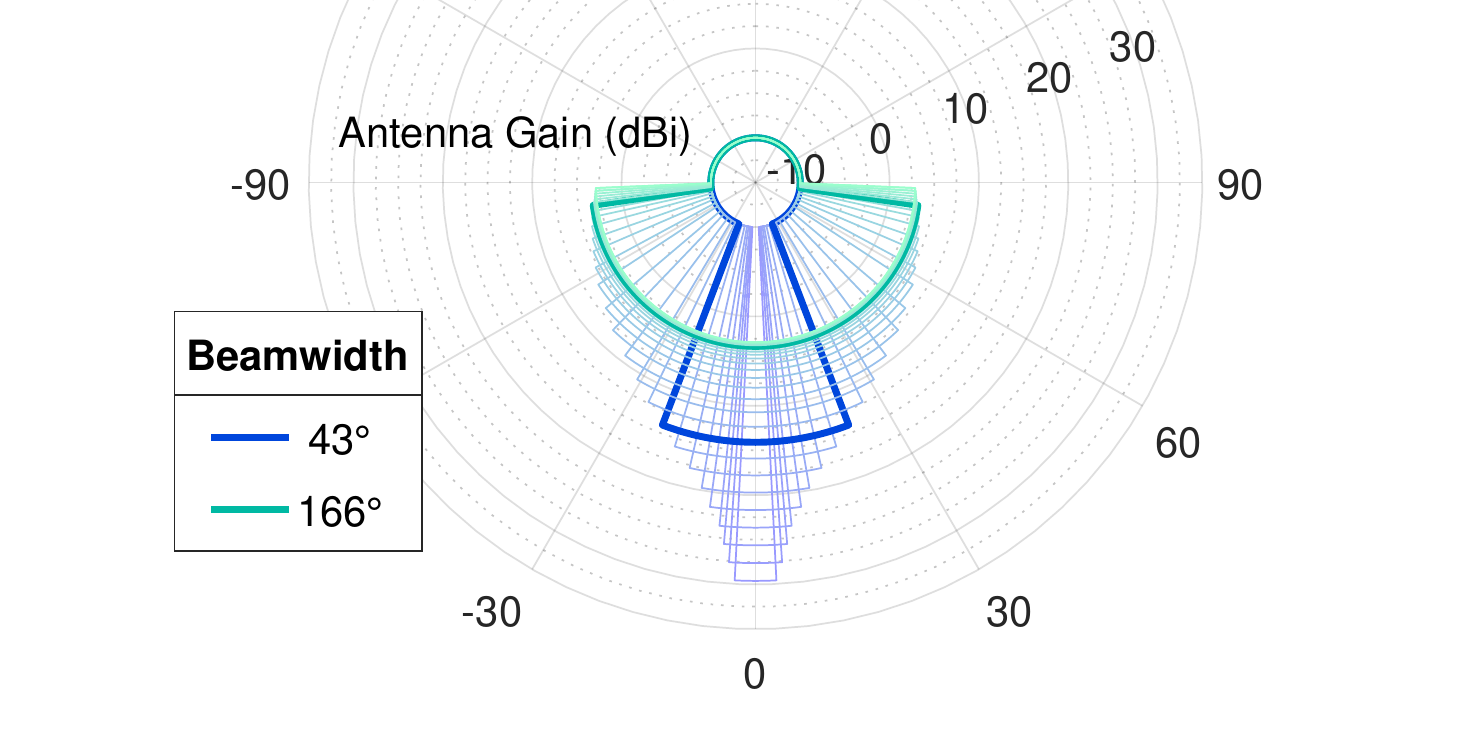}
        \caption{Antenna directivity gain pattern for different beamwidths, according to (\ref{eq:mainlobegain}).}
        \label{fig:antennapattern}
    \end{subfigure}
    \caption{Directional antenna model. }
\end{figure}

We assume the APs utilise fixed directional transmission due to its advantage of simplicity.
It does not require phase shifters, which usually require complex processing for precise adjustments \cite{itu-r_m.2376-0}.
We assume the UEs can steer the directional reception, which is perfectly aligned to the direction of the serving AP, 
since there are efficient beam search techniques for static devices within short distances in mmWave systems \cite{ku2016efficient}.
We assume the antenna pattern follows the \yale{cone-bulb} model,
which allocates the most significant part of the signal's energy to the main-lobe, as illustrated in Fig.~\ref{fig:conebulbmodel}, while conserving the total energy of the propagating signal \cite{ramanathan2001performance}.
The \yale{cone} represents the main-lobe attached to a single \yale{bulb} representing the side-lobe. The main-lobe directivity gain is given by $m$, $s$ is the side-lobe gain, $\omega_\mathrm{A}$ is the beamwidth of the main-lobe, $\mathrm{cap}$ is the area of the spherical cap, and $\mathrm{sph}$ is the surface area of the sphere of radius $r$.
The directivity gain is then a function of the beamwidth, normalised over a given spherical surface as in:
\begin{equation}
    m \cdot \frac{\mathrm{cap}}{\mathrm{sph}} + s \cdot \frac{\mathrm{sph}-\mathrm{cap}}{\mathrm{sph}} = 1,
    \label{eq:beamformgain}
\end{equation}
where $\mathrm{cap} = 2 \pi r^2 \left(1-\cos \frac{\omega_\mathrm{A}}{2}\right)$, and $\mathrm{sph} = 4 \pi r^2$. Thus, fixing the side-lobe gain $s$, we can calculate the main-lobe gain as a function of the beamwidth, as illustrated in Fig.~\ref{fig:antennapattern}:
\begin{equation}
   m = \frac{2 - s \left( 1 + \cos \frac{\omega_\mathrm{A}}{2} \right)}{1 - \cos \frac{\omega_\mathrm{A}}{2}}.
    \label{eq:mainlobegain}
\end{equation}

\begin{figure}[t]
    \centering
    \includegraphics[width=.8\linewidth]{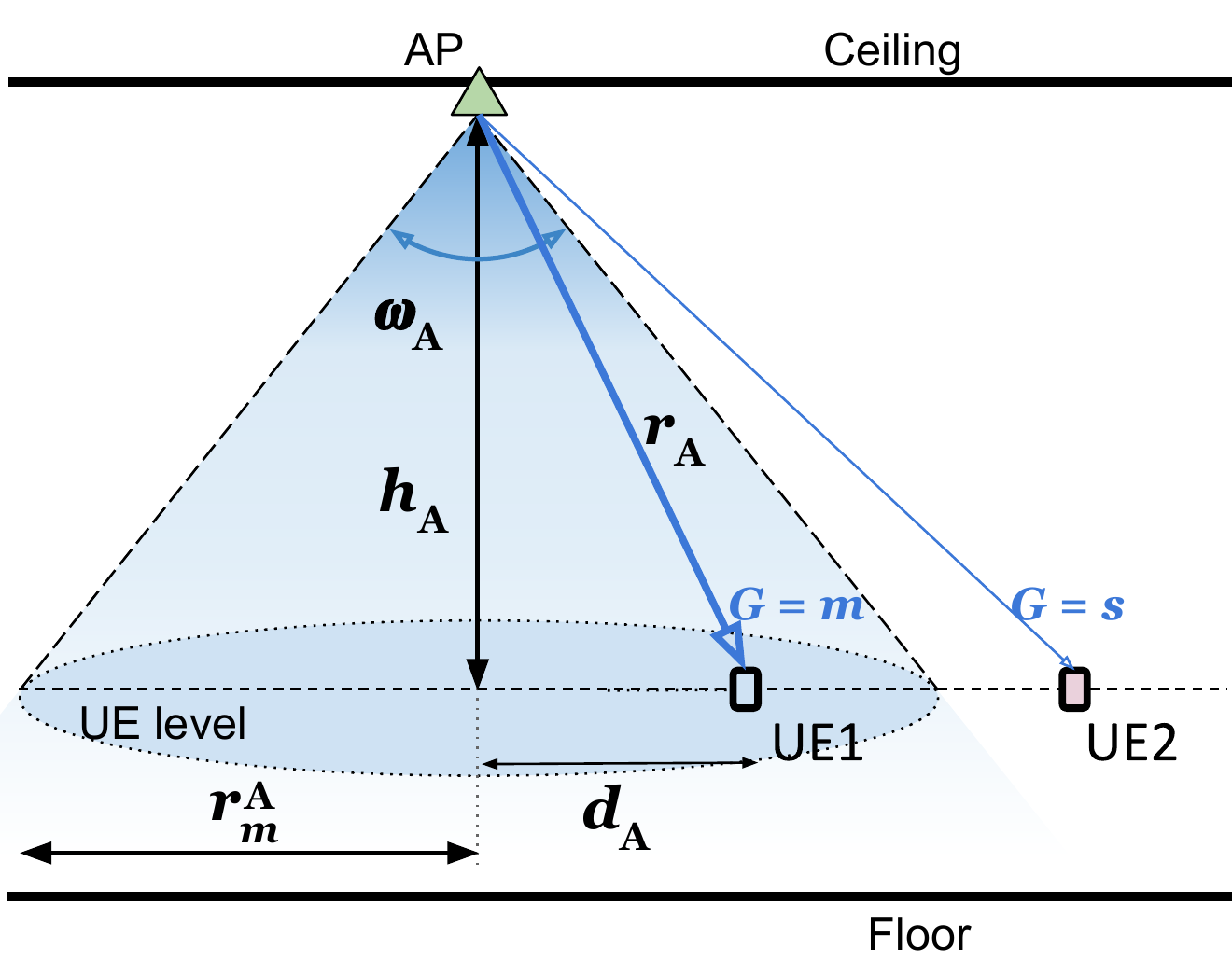}
    \caption{AP directional transmission with beamwidth $\omega_\mathrm{A}$ illuminating a circle with radius $r_m$. The received signal power of UE1 inside the circle has a directivity gain $m$ from the AP at an Euclidean distance $r_\mathrm{A}$, while the received signal power of UE2 outside the circle is scaled by $s$.}
    \label{fig:ap_mainlobe_illumination}
\end{figure}

{ \subsubsection{AP Directivity Gain}}
The \ac{UE} has the maximum transmit directivity gain $m$ when the \ac{UE} is positioned under the area illuminated by the \ac{AP}'s main-lobe, i.e., the \ac{UE} is inside the projected circle of radius:
\begin{equation}
    r_m^\mathrm{A} = h_\mathrm{A} \cdot \tan\frac{\omega_\mathrm{A}}{2},
    \label{eq:ap_mainloberadius}
\end{equation}
as illustrated in Fig.~\ref{fig:ap_mainlobe_illumination}.
Otherwise, the transmit directivity gain is the side-lobe gain $s$.
Therefore, we can express the transmit directivity gain as:
{
\begin{equation}
G^\mathrm{A} = \left\{ \;
\begin{array}{ll}
    m, & d_\mathrm{A} \leq r_m^\mathrm{A};\\
    s, &\text{otherwise};
\end{array}
\; \right.
\label{eq:ap_directivitygain}
\end{equation}}
where $d_\mathrm{A}$ is the projection of the distance from the AP to the UE onto the horizontal plane.

{ 
\subsubsection{UE Directivity Gain}
The signal has maximum receive gain $m$ when the UE's main-lobe illuminates the transmit AP.
We consider the UE's main-lobe is pointed towards the serving AP.
Thus, an AP is illuminated by the UE's main-lobe cone when the AP is located in the conic\footnote{ In mathematics, a conic section (or simply conic) is a curve obtained as the intersection of the surface of a cone with a plane.} shape projected around the serving AP on the ceiling surface, as illustrated in Fig.~\ref{fig:ue_mainlobe_projection}.
\begin{figure}
    \centering
    \begin{subfigure}{.4\linewidth}
        \centering
    \includegraphics[width=\linewidth]{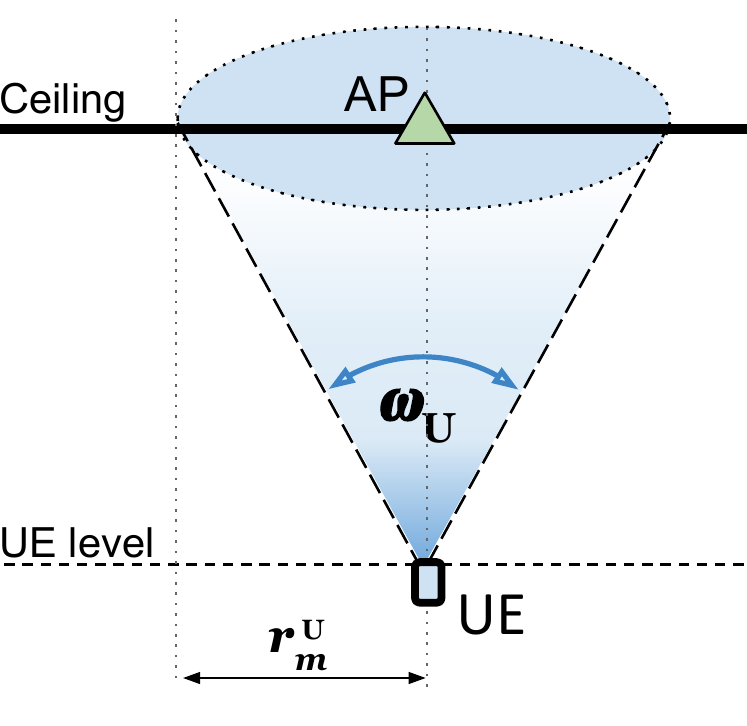}
    \caption{Bounded illumination area is a circle.}
    \label{fig:ue_mainlobe_illumination_bounded}
    \end{subfigure}%
    \hspace{.3cm}
    \begin{subfigure}{.48\linewidth}
        \centering
    \includegraphics[width=\linewidth]{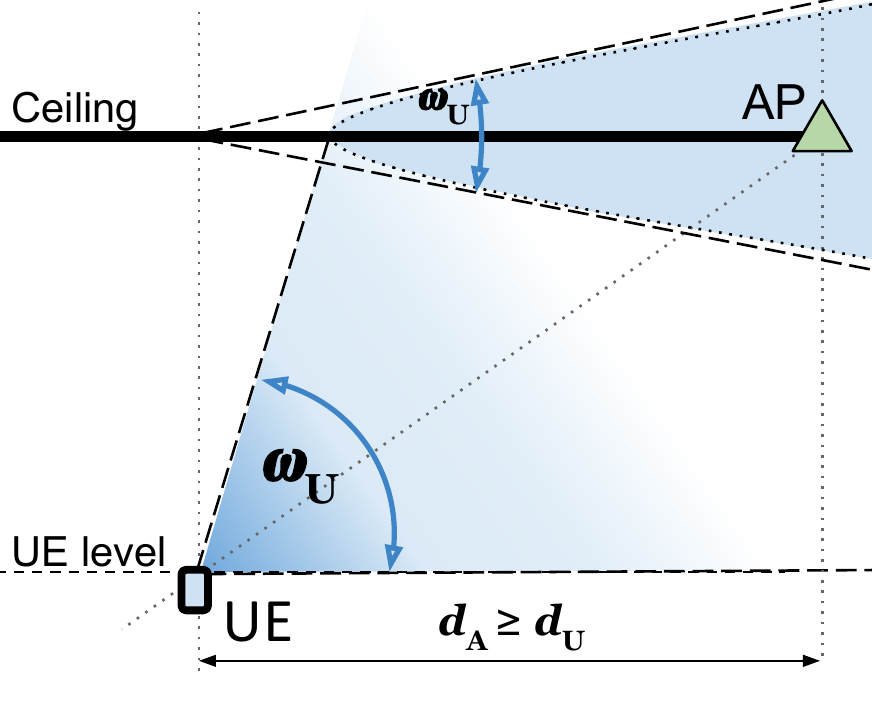}
    \caption{Unbounded illumination area is a circular sector$^2$.}
    \label{fig:ue_mainlobe_illumination_unbounded}
    \end{subfigure}
    \caption{UE directional reception with beamwidth $\omega_\mathrm{U}$ illuminating a conic section on the ceiling surface. The received signal power from an AP inside the cone has a directivity gain $m$, while the received signal power from an AP outside the cone is scaled by $s$.}
    \label{fig:ue_mainlobe_projection}
\end{figure}
As the UE gets further apart from the serving AP, the UE's main-lobe elevation lowers, and eventually, the lateral surface of the cone becomes parallel to the UE level, making the illuminated area in the ceiling to transit from bounded to unbounded. This event happens at a distance of $d_\mathrm{U}$:
\begin{equation}
    d_\mathrm{U} = \frac{h_\mathrm{A}}{\tan\frac{\omega_\mathrm{U}}{2}}.
    \label{eq:ue_mainlobe_projection_bound}
\end{equation}
Hence, we assume that, if the UE is close to the serving AP, i.e., $d_\mathrm{A}<d_\mathrm{U}$, the illuminated area is bounded and the conic shape is a circle of radius $r_m^\mathrm{U}$:
\begin{equation}
    r_m^\mathrm{U} = h_\mathrm{A} \cdot \tan\frac{\omega_\mathrm{U}}{2},
    \label{eq:ue_mainloberadius}
\end{equation}
If the UE is far enough from the serving AP, i.e., $d_\mathrm{A} \geq d_\mathrm{U}$, the conic projected in the ceiling becomes unbounded, thus we assume that the illuminated area is a circular sector\footnote{ A circular sector is the portion of a disk enclosed by two radii and an arc. The length of the radii is limited by the deployment area. \label{fn:circular_sector}} with central angle equal to the UE beamwidth, as illustrated in Fig.~\ref{fig:ue_mainlobe_illumination_unbounded}.
Thus, since the UE's directionality is aligned with the serving AP orientation, an AP is illuminated by the UE's main-lobe when the AP orientation falls within the illuminated area around the serving AP.
Therefore, we can express the receive directivity gain as:
\begin{equation}
G^\mathrm{U} = \left\{
\begin{array}{ll}
    m, & \left(d_\mathrm{A}<d_\mathrm{U} \;\; \cap \;\; d_\mathrm{A} \leq r_m^\mathrm{U}\right) \; \cup \; \\
    & \big(d_\mathrm{A}\geq d_\mathrm{U} \;\;\; \cap  \\
    & \; \theta_\mathrm{S} - \frac{\omega_\mathrm{U}}{2} < \theta_\mathrm{A} < \theta_\mathrm{S} + \frac{\omega_\mathrm{U}}{2}\big); \\
    s, & \text{otherwise};
\end{array}
\; \right.
\label{eq:ue_directivitygain}
\end{equation}
where $\theta_\mathrm{A}$ is the orientation of the given AP and $\theta_\mathrm{S}$ is the serving AP orientation.
}

\subsection{Body Blockage}
\label{sec:bodyblockage}

\begin{figure}
   \centering 
    \begin{subfigure}{\linewidth}
        \centering
        \includegraphics[width=.8\linewidth]{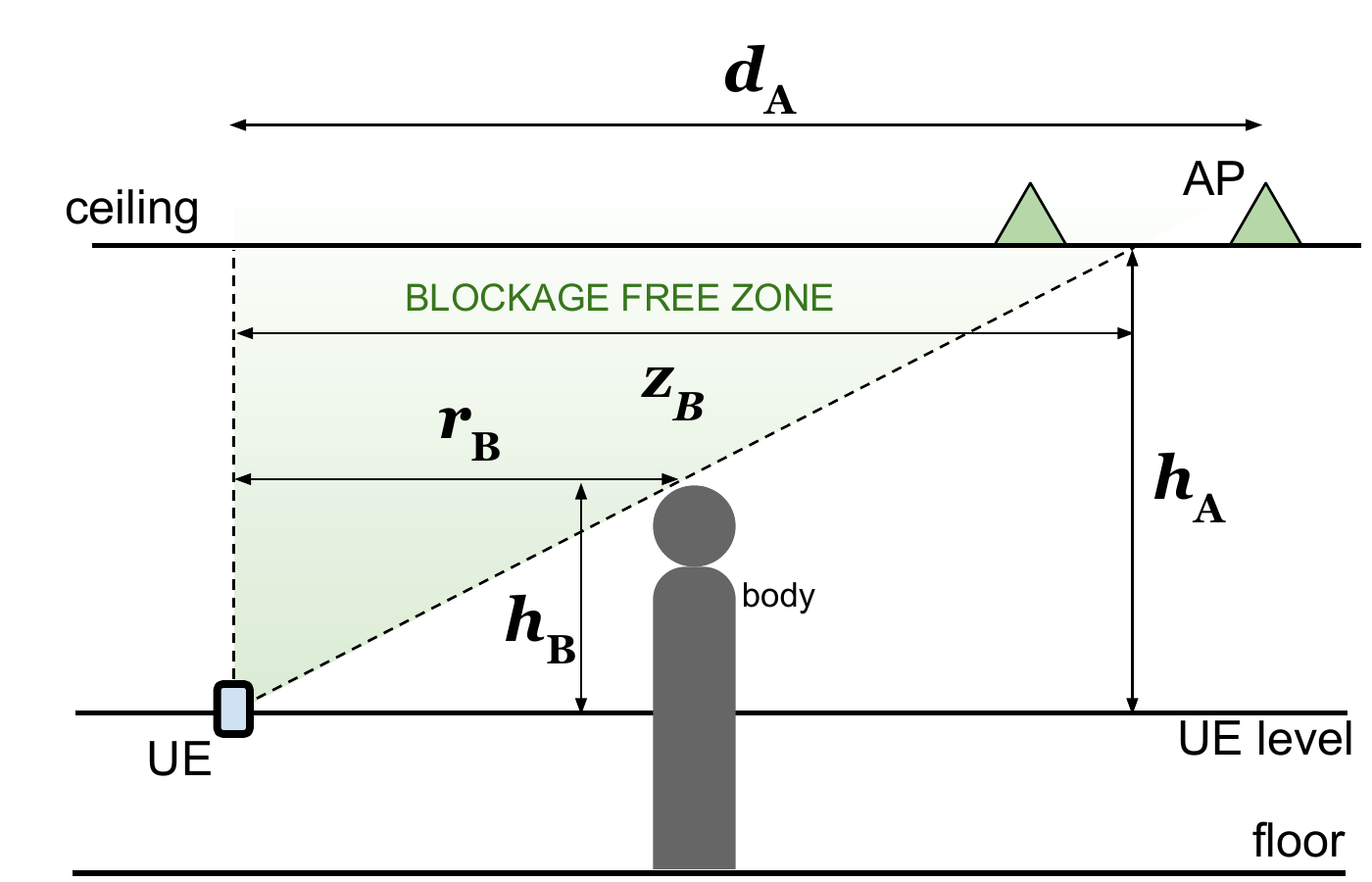}
        \caption{Side view. Obstruction by a body may happen when the \ac{AP} is beyond the radius $z_\mathrm{B}$. On the other hand, an \ac{AP} within this radius is not obstructed, regardless the body orientation with respect to the \ac{UE}, i.e., the \ac{AP} is inside the \textit{blockage free zone} represented by the green area.}
        \label{fig:body_blockage_side_view}
    \end{subfigure}
    \hfill
    \begin{subfigure}[t]{.48\linewidth}
        \centering
        \includegraphics[width=\linewidth]{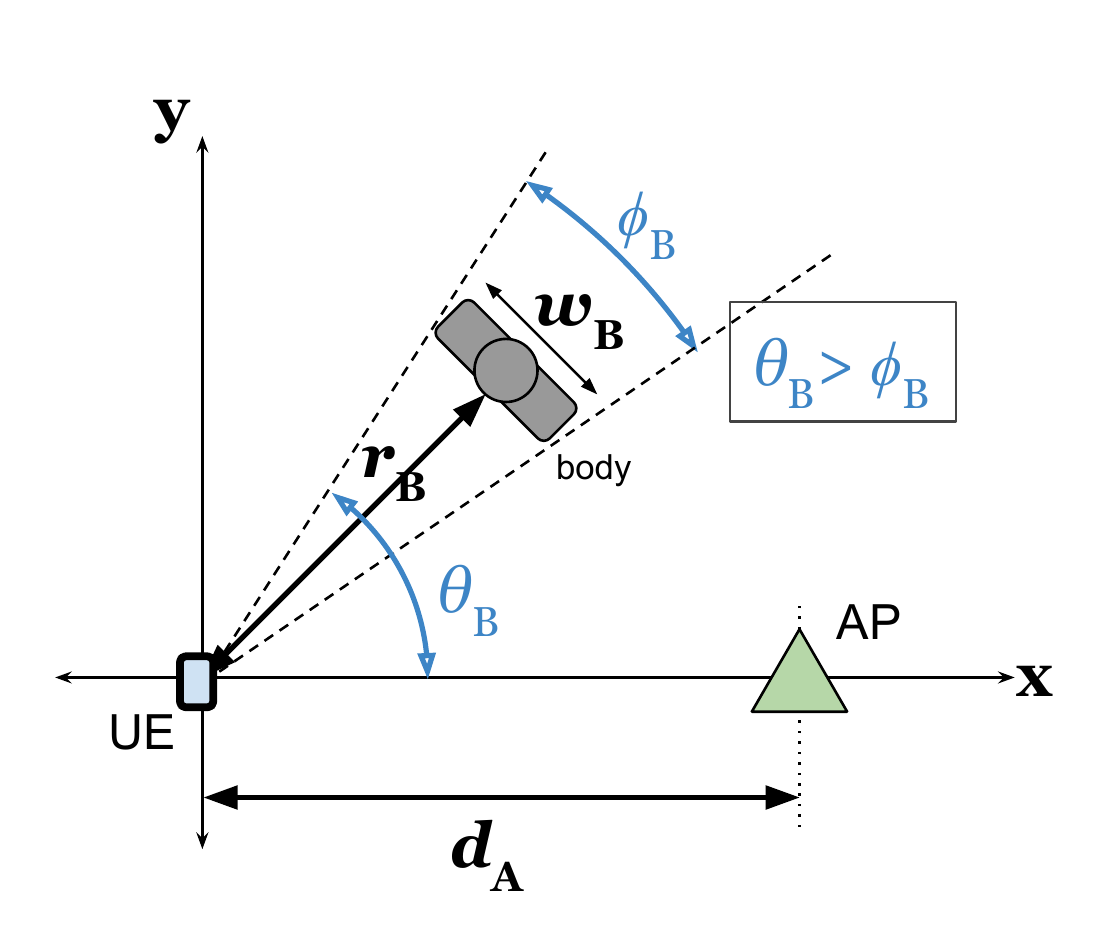}
        \caption{Top view. The body is not between the \ac{AP} and the \ac{UE} when the shadowed angle $\phi_\mathrm{B}$ is less than the angle $\theta_\mathrm{B}$.}
        \label{fig:body_blocking_angle}
    \end{subfigure}%
    \hfill
    \begin{subfigure}[t]{.48\linewidth}
        \centering
        \includegraphics[width=\linewidth]{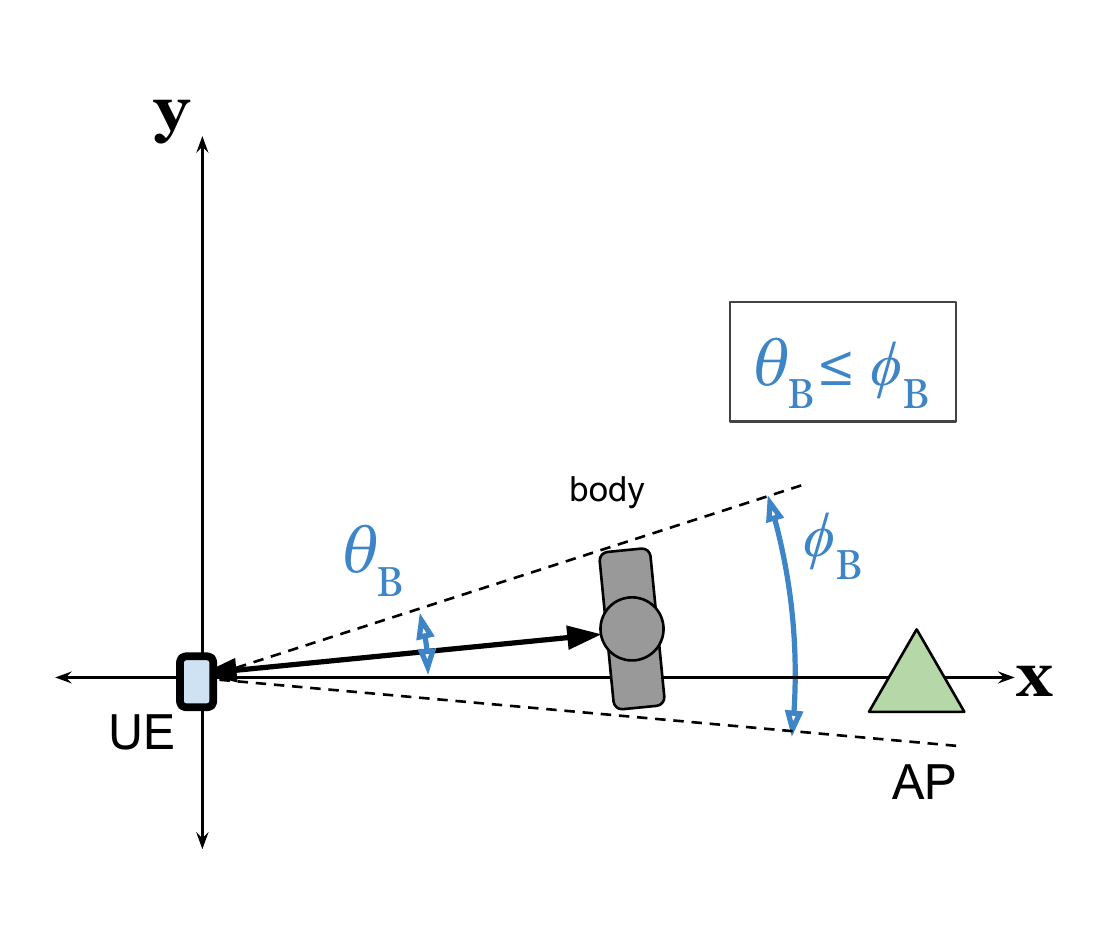}
        \caption{Top view. The body is between the \ac{AP} and the \ac{UE}, and possibly blocking the \ac{AP}, when the shadowed angle $\phi_\mathrm{B}$ is greater than the angle $\theta_\mathrm{B}$.}
        \label{fig:body_blocking_angle_blocked}
    \end{subfigure}
    \caption{Body blockage model. The body is at a distance of $r_\mathrm{B}$ from \ac{UE}, and height $h_\mathrm{B}$ from UE level, and has a width $w_\mathrm{B}$.}
    \label{fig:bodyblock}
\end{figure}

In this model, the main factor that describes 
how likely it is for a human body to shadow signals to/from the \ac{UE} 
is the \ac{UE}'s position with respect to the body. 
{ We assume that a body facing\footnote{Without loss of generality, we assume that all bodies, not only the user's body, are facing the UE. The effect of having a body not perpendicular to the UE, i.e. smaller $w_\mathrm{B}$, is to have a smaller blocking angle, which is equivalent to the body being at a longer distance $r_\mathrm{B}$.} the UE shadows an area given by a rectangle of height $h_\mathrm{B}$ and width $w_\mathrm{B}$. The UE is in front of the body at a distance $r_\mathrm{B}$ from the body centre as shown in Fig.~\ref{fig:bodyblock}.}
In the horizontal plane, the obstructed space can be quantified by its angle (shadowing angle) $\phi_\mathrm{B} = 2 \arctan ({w_\mathrm{B}}/{2 r_\mathrm{B}})$.
The parameter $h_\mathrm{B}$ determines the space obstructed by the body in the vertical plane.

We assume that the signal from an \ac{AP} can be obstructed by the \ac{UB} (the body holding the \ac{UE}) or by other \acp{RB}. We define these events as \textit{self-body blockage} and \textit{random-body blockage}, respectively.
We assume that all bodies (\ac{UB} and \acp{RB}) have the same size ($w_\mathrm{B}$ and $h_\mathrm{B}$ are constant). The \ac{UB} is fixed at distance $r_0$ from the \ac{UE}, while the \acp{RB} are uniformly placed, thus the distance $r_\mathrm{B}$ from the UE to an \ac{RB} is random. 
We define the orientation $\theta$ with respect to the body as the angle between the line joining the UE to the body's right shoulder of the body and the line joining the UE to the AP. Then, we assume that the UB and RB orientations, $\theta_0$ and $\theta_\mathrm{B}$ angles respectively, are uniformly distributed in $[0,2\pi)$.

Given the body blockage and our ceiling-mounted deployments, we can construct a model of user device shadowing as depicted in Fig.~\ref{fig:bodyblock}. From the geometry of the model, and assuming all APs at the same height $h_\mathrm{A}$, we define $z_\mathrm{B}$ as the radius of the \textit{blockage free zone} for a given \ac{UE}-body pair (illustrated in Fig.~\ref{fig:body_blockage_side_view}):
\begin{equation}
    z_\mathrm{B} = h_\mathrm{A} \cdot \frac{r_B}{h_B},
    \label{eq:blockfreeradius}
\end{equation}
where an \ac{AP} inside this zone will always be in \ac{LOS} with the \ac{UE}, regardless of the body orientation.
The \acp{AP} outside this zone will have the \ac{LOS} obstructed whenever the body is in between the \ac{UE} and the \ac{AP} (considering their projections onto the two-dimensional plane).
{ For a given distance between UE and AP, and a given distance between UE and body blockage}, and $\mathbbm{1}(.)$ denoting the indicator function, we define the following events: 
\begin{definition}
\label{def:ApOutside}
The event \textit{'AP outside'} = \textit{[the \ac{AP} is outside the blockage free zone]}, as illustrated in Fig.~\ref{fig:body_blockage_side_view}, occurs when the distance $d_\mathrm{A}$ from the AP to the UE is greater than the blockage free zone radius $z_\mathrm{B}$, i.e.: 
\begin{equation}
    \mathsf{P}(\text{AP outside})=\mathbbm{1}(d_\mathrm{A} > z_\mathrm{B}).
\end{equation}
\end{definition}

\begin{definition}
\label{def:bodyBetween}
The event \textit{'body between'} = \textit{[the body is in between the \ac{AP} and the \ac{UE}]}, as illustrated in Fig.~\ref{fig:body_blocking_angle_blocked}, occurs when the body orientation angle $\theta_\mathrm{B} \in [0,2\pi)$ is smaller than the body shadowing angle $\phi_\mathrm{B} \in [0,\pi)$, i.e.:
\begin{equation}
    \mathsf{P}(\text{body between})=\mathbbm{1}(\theta_\mathrm{B} < \phi_\mathrm{B}).
\end{equation}
Note that the UB shadowing angle is constant since the UB is at a fixed distance from the UE. The RB shadowing angle is a random variable since the distance from the UE is random, and its distribution is derived in Appendix \ref{ap:distributionPhi}.
\end{definition}
\begin{definition}
The event \textit{'AP blocked'} = \textit{[the \ac{AP} is blocked by the body]} is the intersection of the events \textit{[the \ac{AP} is outside the blockage free zone]} \textbf{and} \textit{[the body is in between the \ac{AP} and the \ac{UE}]}, i.e.:
\begin{equation}
\begin{split}
    \mathsf{P}(\text{AP blocked}) &= \mathsf{P}(\text{AP outside} \; \cap \; \text{body between}) \\
    &= \mathbbm{1}(d_\mathrm{A} > z_\mathrm{B} \; \cap \; \theta_\mathrm{B} < \phi_\mathrm{B}).
\end{split}
\end{equation}
\label{def:bodyblockage}
\end{definition}

Then, averaging over all the random locations of the body blockages, including the user body, and the distance between the UE and an AP, we define the following for a given pair UE-AP:
\begin{definition}
\label{def:ApBlocked}
Assuming that one body is enough to cause signal blockage, the event \textit{[AP being blocked by a body (UB or RB)]} can be expressed as \textit{[at least one RB is blocking]} \textbf{or} \textit{[UB is blocking]}. We define the probability  $p_\mathrm{A}$ of this event as: 
\begin{equation}
\begin{split}
    p_\mathrm{A} &= \mathsf{P}(\text{at least one random body is blocking} \\
    &\;\;\;\;\; \cup \text{ user body is blocking}).
\end{split}
\label{eq:defpa}
\end{equation}
\end{definition}

{ 
\begin{definition}
\label{def:blockageState} We define the blockage state as $X$ which can take values from the set $\chi = \mbox{\footnotesize $\{\mathrm{LOS},\mathrm{NLOS}\}$}$, where the {\footnotesize $\mathrm{NLOS}$} is given by the event \textit{[AP being blocked by a body (UB or RB)]}, with the {\footnotesize $\mathrm{LOS}$} being the complementary event. Thus, we express $X$ as:
\begin{equation}
     X = \left\{ \;
\begin{array}{lll}
    \mbox{\footnotesize $\mathrm{LOS}$}, & \text{w.p.} & 1-p_\mathrm{A}; \\
    \mbox{\footnotesize $\mathrm{NLOS}$}, & \text{w.p.} & p_\mathrm{A};
\end{array}
\; \right.
\end{equation}
\end{definition}
}

\subsection{Signal-to-Interference-plus-Noise Ratio}
\label{sec:sinr}
In this work, we make the following assumptions to compute the SINR:
\begin{assumption}
\label{assume:independentblockage}
We assume that the blockage event for each \ac{AP} is independent, i.e., there is no correlation in the blockage process of \acp{AP} near each other, even though in reality they could be blocked by the same body \cite{venugopal2016device}. Therefore, we can determine the probability of blockage as a function of the AP distance. In Section \ref{sec:analyticalresults}, we test this assumption and check whether the correlation in body blockage has any impact on the network performance.
\end{assumption}

\begin{assumption}
{
We model the mmWave signal propagation considering the experimentally-validated channel model proposed in \cite{yoo2017measurements}.
The model is composed of the path loss, large-scale fading (shadowing) and small-scale fading, whose parameters assume one of two values according to the blockage state $X$.
We assume the path loss conditioned on $X=x$ is:
\begin{equation}
    L_x = \ell_x \cdot r_\mathrm{A}^{-\nu_x},
    \label{eq:los.pathloss}
\end{equation}
where $\ell_x$ is the path loss at 1 metre distance under free space propagation, $\nu_x$ is the attenuation exponent, $r_\mathrm{A} = \sqrt{d_\mathrm{A}^2+h_\mathrm{A}^2}$ is the Euclidean distance from the \ac{AP} to the \ac{UE}, and $d_\mathrm{A}$ is the projection of the distance from the cell centre to the \ac{UE} onto the horizontal plane (we refer to this as 2D-distance, as shown in Fig.~\ref{fig:ap_mainlobe_illumination}).
We assume the shadowing is modelled as Gamma distribution and the small-scale fading as Nakagami-m.}
\end{assumption}

{
We express the received power at the UE from an AP with blockage state $X=x$ as:
\begin{equation}
    P_{\mathrm{r},x} =  p_\mathrm{t} \cdot G^\mathrm{A}_x \cdot G^\mathrm{U}_x \cdot L_x \cdot B_x \cdot H_x,
\end{equation}
where $ p_\mathrm{t}$ is the transmit power, $G^\mathrm{A}_x$ is the AP directivity gain given in (\ref{eq:ap_directivitygain}), $G^\mathrm{U}_x$ is the UE directivity gain given in (\ref{eq:ue_directivitygain}), $L_x$ is the path loss given in (\ref{eq:los.pathloss}), $B_x$ is the shadowing gain, and $H_x$ is the small-scale fading gain.

Based on the assumptions made above, we can express the SINR at a \ac{UE} as follows :
\begin{equation}
    \mathrm{SINR} = \frac{G^\mathrm{A}_{x_i} \cdot G^\mathrm{U}_{x_i} \cdot L_{x_i} \cdot B_{x_i} \cdot H_{x_i}}{ \frac{\sigma}{p_\mathrm{t}} + \sum_{j \in \Ac \setminus \{i\}} G^\mathrm{A}_{x_j} \cdot G^\mathrm{U}_{x_j} \cdot L_{x_j} \cdot B_{x_j} \cdot H_{x_j}},
\end{equation}
where $i\in\Ac$ is the serving AP, with $\Ac$ denoting the set of APs, and $\sigma$ is the thermal noise power. 
Note that, because $X$ is a random event, $G^\mathrm{A}_{x}$, $G^\mathrm{U}_{x}$, $L_x$ are random variables whose distributions are functions of the system parameters ($\omega_\mathrm{A}$, $\omega_\mathrm{U}$, $h_\mathrm{A}$, $r_B$, $h_B$, $w_B$). A summary of all the system deterministic parameters and random variables is provided in Table~\ref{tab:parameters} and in Table~\ref{tab:variables}, respectively.
}
\begin{table}[b]
\centering
\caption{Summary of the System Parameters}
\label{tab:parameters}
\begin{tabular}{@{}llll@{}}
\toprule
\multicolumn{2}{c}{\textbf{Body Parameters}}    & \multicolumn{2}{c}{\textbf{Deployment Parameters}} \\ 
\midrule
Body width          & $w_\mathrm{B}$      & \ac{AP} height      & $h_\mathrm{A}$      \\
Body height         & $h_\mathrm{B}$      & \ac{AP} beamwidth   & $\omega_\mathrm{A}$ \\
Distance from \ac{UE} to \ac{UB}  & $r_0$ & Inter-site distance & $\delta$    \\
 &  & \ac{UE} beamwidth   & $\omega_\mathrm{U}$ \\ \midrule
\multicolumn{2}{c}{\textbf{Signal Power Parameters}} & \multicolumn{2}{c}{\textbf{Directivity Gain Parameters}} \\
\midrule
Noise power         & $\sigma$             & Main-lobe gain      & $m$   \\
Transmit power      & $p_\mathrm{t}$ & Side-lobe gain      & $s$   \\
                    &                 & Illumination radius & $r_m$ \\ 
\bottomrule
\end{tabular}
\end{table}

\begin{table}[]
\centering
\caption{Summary of the System Random Variables}
\label{tab:variables}
\begin{tabular}{@{}llll@{}}
\toprule
\multicolumn{2}{c}{\textbf{Body Blockage Variables}}    & \multicolumn{2}{c}{\textbf{Signal Power Variables}} \\ 
\midrule
Body shadowing angle              & $\phi_\mathrm{B}$   & Directivity gain  & $G$   \\
Body orientation angle            & $\theta_\mathrm{B}$ & Path gain         & $L$   \\
Distance from \ac{UE} to \ac{RB}  & $r_\mathrm{B}$      & Shadowing gain     & $B$   \\
Blockage free zone radius                & $z_\mathrm{B}$      & Small-scale fading gain       & $H$   \\
                                  &                     & Received power    & $P_\mathrm{r}$ \\
\midrule
\multicolumn{4}{c}{\textbf{Deployment Variables}} \\
\midrule
Distance from \ac{UE} to \ac{AP}  & $r_\mathrm{A}$      & 2D projection of $r_\mathrm{A}$ & $d_\mathrm{A}$ \\
\bottomrule
\end{tabular}
\end{table}

\section{Analytical Results}
\label{sec:analyticalresults}
In this Section, we provide an analytical expression for the probability of an AP being blocked. We use this expression to reduce the complexity of system-level simulation.\footnote{The simulator would need to compute the inequalities in Definitions \ref{def:ApOutside} and \ref{def:bodyBetween} for each AP-body pair, i.e., $n_\mathrm{A} \times n_\mathrm{B}$ times (number of APs and number of RBs, respectively). With the analytical expression, the simulator only needs to calculate the function in (\ref{eq:defpa}) for each AP, i.e., $n_A$ times.}
In addition, we validate the \ac{AP} blockage event independence assumption
{ by showing that the proposed analytical expressions yield results closely trailing numerically evaluated blockage probabilities obtained from our simulation environment.}

To compute the blockage of a signal from an \ac{AP} at a given distance $d_\mathrm{A}$ from the \ac{UE}, we need to find the expression that gives the probability $p_\mathrm{A}$ as a function of the distance. This function depends on two events --- the \textit{self-body blockage} and the \textit{random-body blockage} --- and on the number of random bodies in the venue.
{ Assuming that the RBs are randomly placed in a square-shaped venue, and the blockage orientation is distributed uniformly, we state the following propositions:}

\begin{proposition}
The probability of self-body blockage is given by
\begin{equation}
    p_0(d_\mathrm{A}) = \left\{\;
    \begin{split}
        & \frac{\arctan(w_\mathrm{B}/_{2r_0})}{\pi}, \; d_\mathrm{A} \geq r_0\frac{h_\mathrm{A}}{h_\mathrm{B}}; \\
        & 0,\;\; 0 < d_\mathrm{A} < r_0 \frac{h_\mathrm{A}}{h_\mathrm{B}};
    \end{split}
    \; \right.
\label{eq:probSelfBodyBlockage}
\end{equation}
\end{proposition}
\begin{IEEEproof}
See the proof in Appendix \ref{ap:proofp0}.
\end{IEEEproof}

\begin{proposition}
The probability of blockage by a random body is given by
\begin{align}
    p_1(d_\mathrm{A}) &= \int\limits_{\varphi( d_\mathrm{A} \frac{h_\mathrm{B}}{h_\mathrm{A}} )}^\pi \frac{-w_\mathrm{B} \; \phi}{2\pi(\cos\phi-1)} \left( \frac{\pi \rho}{s^2} - \frac{4 \rho^2}{s^3} + \frac{\rho^3}{s^4}  \right) \; \mathrm{d}\phi,
\label{eq:probApBlockedByOneBody}
\end{align}
where $\varphi(x)$ is the angle of obstruction by the body as a function of a distance $x$, $\rho = w_\mathrm{B}/(2\tan\frac{\phi}{2})$ and $s$ is the square side length.
\end{proposition}
\begin{IEEEproof}
See the proof in Appendix \ref{ap:proofp1}.
\end{IEEEproof}

\begin{proposition}
Considering the probabilities in (\ref{eq:probApBlockedByOneBody}) and (\ref{eq:probSelfBodyBlockage}), and a finite number $N_\mathrm{B}$ of \acp{RB} in the venue, we express the probability of an \ac{AP} at a distance $d_\mathrm{A}$ from the \ac{UE} being blocked by a body (\ac{UB} or a \ac{RB}) as
\begin{equation}
\begin{split}
    p_\mathrm{A}(d_\mathrm{A})
    &= 1 - \big(1-p_1(d_\mathrm{A})\big)^{N_\mathrm{B}} \times \big(1- p_0(d_\mathrm{A})\big).
\end{split}
\label{eq:probApBlocked}
\end{equation}
\end{proposition}
\begin{IEEEproof}
Applying the Definition \ref{def:ApBlocked}, we can express the probability as:
\begin{equation}
\begin{split}
    p_\mathrm{A}(d_\mathrm{A}) &= \mathsf{P}(\text{at least one random body is blocking} \\
    &\;\;\;\;\; \text{\textbf{or} user body is blocking}) \\
    &= 1 - \mathsf{P}(\text{no random body is blocking} \\ 
    &\;\;\;\;\; \text{\textbf{and}  user body is not blocking}) \\
    &= 1 -  \mathsf{P}(\text{no random body is blocking}) \\
    &\;\;\;\;\; \times \mathsf{P}(\text{user body is not blocking}) \\
    &= 1 - \big(1-p_1(d_\mathrm{A})\big)^{N_\mathrm{B}} \times \big(1- p_0(d_\mathrm{A})\big).
\end{split}
\end{equation}
\end{IEEEproof}

\begin{figure}[h!]
    \includegraphics[width=\linewidth,trim={0 2.2cm 0 2.9cm},clip]{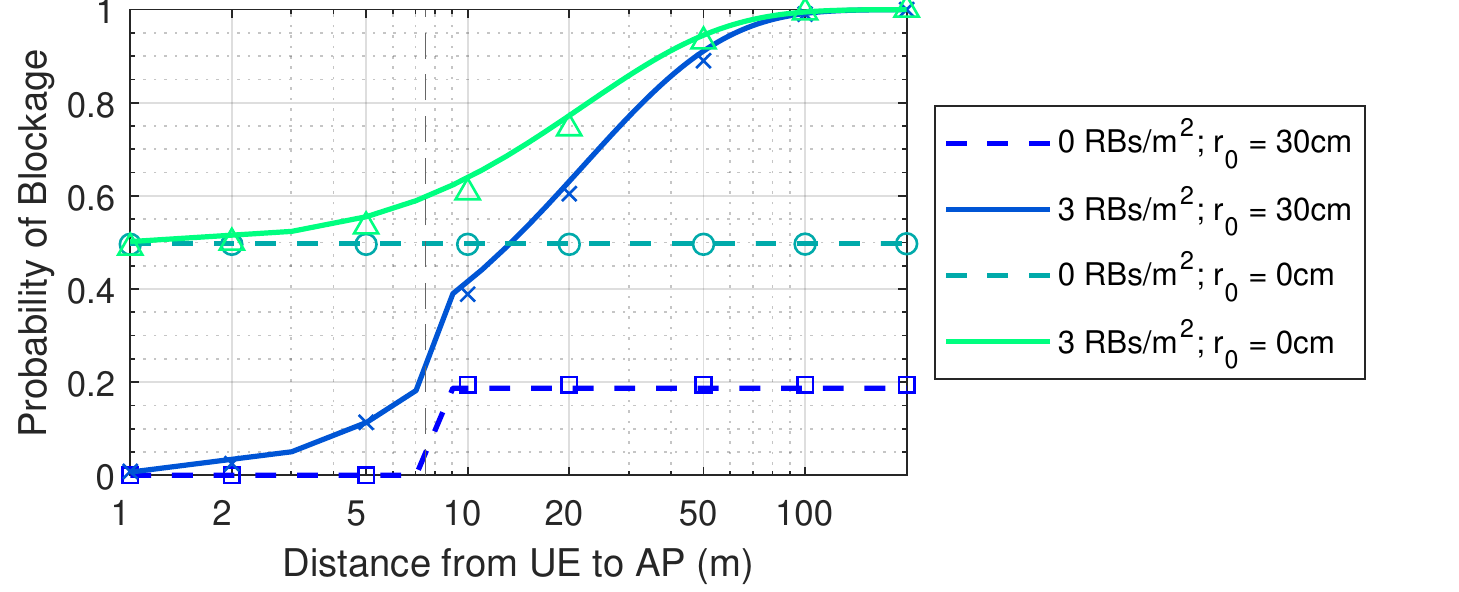}
    
    \caption{Analytical probability of Body Blockage $p_\mathrm{A}(d)$ (lines) and relative frequency of blockages in Monte-Carlo simulations (markers). Comparison between different blockage scenarios. The grey vertical dashed line is the radius of the self-body blockage free zone.}
    \label{fig:probBlockage}
\end{figure}

To validate the \ac{AP} blockage event independence assumption, we compare (\ref{eq:probApBlocked}) to the outcome of Monte-Carlo simulations 
for the same setup, { where the AP blockage events can be correlated (see Fig.~\ref{fig:probBlockage}). 
We show the results by varying the distance $r_0$ between UE and UB and the number $N_\mathrm{B}$ of RBs. 
We consider a square-shaped venue area of \unit[400$\times$400]{m$^2$} and inter-site distance of \unit[20]{m}.
For the Monte-Carlo simulation, we fix the UE at the origin\footnote{ When simulating the UE in other positions (e.g., in a corner and in an edge of the square-shaped venue), we found no significant deviation from the results reported for the centre position. } and we vary the AP position (distance and orientation with respect to the UE). The UB is at a fixed distance from the UE, and its orientation is uniformly distributed. The RBs coordinates are distributed uniformly in both x-axis and y-axis.}
We observe that the expression in (\ref{eq:probApBlocked}) is a good approximation { compared to the relative frequency of blockages computed by simulation}. One should note that when the \ac{AP} is close enough to the \ac{UE} { ($d_\mathrm{A}<$ \unit[5]{m}) and $r_0=$ \unit[30]{cm}}, the probability of blockage is virtually zero, since the \ac{AP} is inside the self-blockage free zone.

\section{Numerical Results}
\label{sec:numericalresults}
\subsection{Simulation Setup}
\label{sec:simulationsetup}
\begin{figure}[h]
    \centering
    \includegraphics[width=\linewidth]{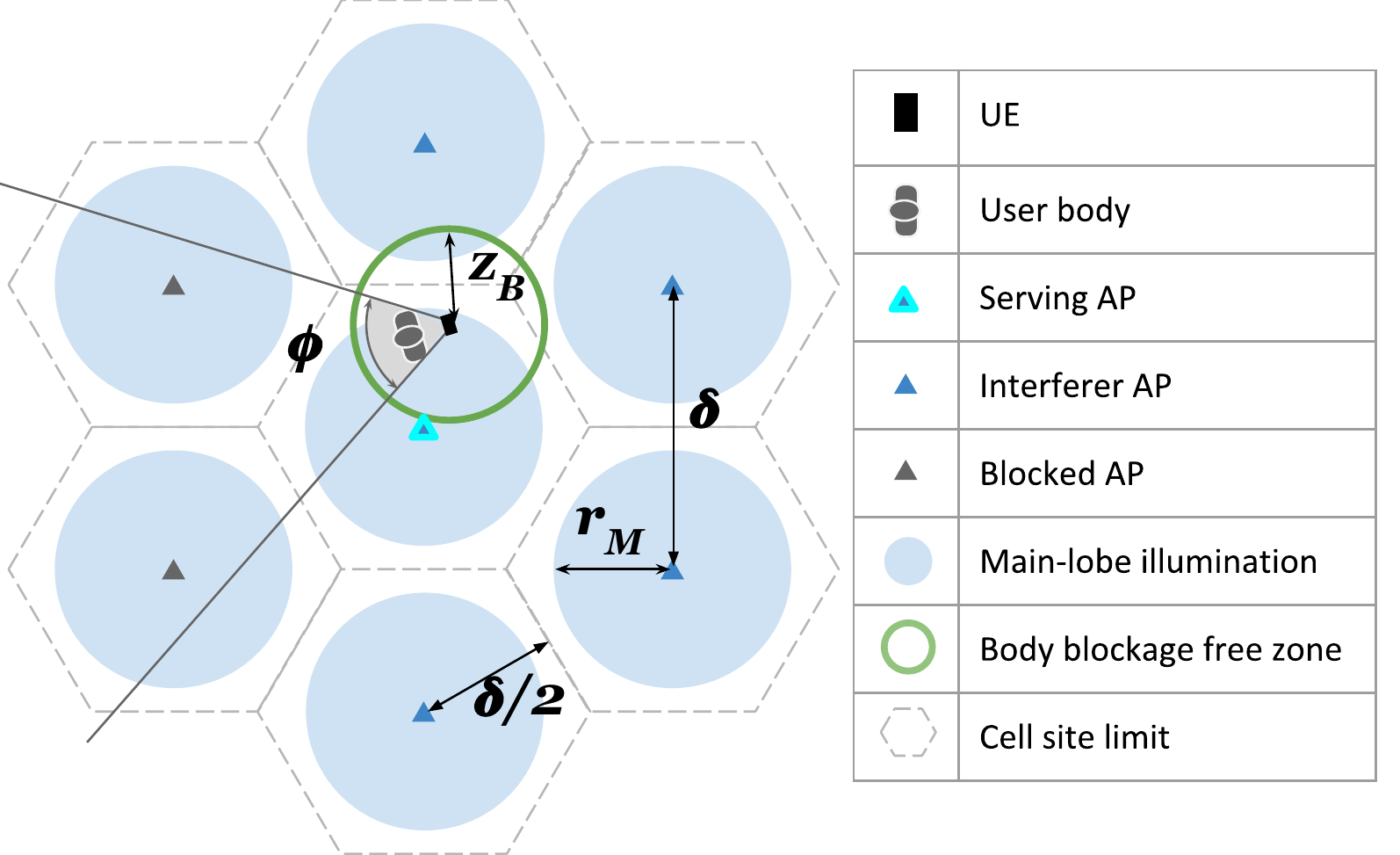}
    \caption{Snapshot from simulations illustrating the system model. The \acp{AP} are distributed according to a hexagonal cell pattern with an inter-site distance $\delta$. The user body is blocking the signal from the grey-coloured \acp{AP} inside the angle $\phi$. The \ac{UE} is illuminated (light-blue area) by the serving \ac{AP}, which is outside the blockage free zone (green circle). Note that in a very dense topology, where $\delta$ could be as small as $z_\mathrm{B}$, there will always be an \ac{AP} (serving or interfering) in \ac{LOS}.}
    \label{fig:topology}
\end{figure}

We model our scenario by placing the \acp{AP} in the centres of a hexagonal cell pattern laid over a \unit[400 $\times$ 400]{m$^2$} area, as exemplified in Fig.~\ref{fig:topology}. 

This specific choice of the area size allows us to explore the system behaviour for large inter-site distances (up to \unit[200]{m}). The side-lobe gain is fixed at \unit[$-10$]{dB}, and the main-lobe gain varies with the beamwidth according to (\ref{eq:mainlobegain}). We evaluate the system for a fixed \ac{AP} height $h_\mathrm{A}$~=~\unit[10]{m}. Note that, changing $h_\mathrm{A}$ has essentially the same impact on the performance as changing the AP beamwidth, since both $h_\mathrm{A}$ and $\omega_\mathrm{A}$ determine the main-lobe illumination area; as a matter of fact, when testing our system for other height values of interest (e.g., heights from 1 to \unit[10]{m}), we observed no significant deviations from the conclusions we present here. 
We set the transmit power as \unit[20]{dBm}, bandwidth as \unit[2]{GHz}, carrier frequency as \unit[60]{GHz}, as recommended by ITU-R \cite{itu-r_f.1497-2}. 
We assume perfect equalisation in the frequency domain as the impact of frequency selectivity can be significantly reduced using effective modulation and equalisation techniques for \unit[60]{GHz} frequencies \cite{daniels2010upclose}.
We set noise figure as \unit[9]{dB}, and the body parameters $w_B$ as \unit[40]{cm} and $h_B$ as \unit[40]{cm}.
We consider the \ac{UE} is associated with an \ac{AP} corresponding to the strongest long-term received signal power, i.e., without considering small-scale fading gain.
We define two self-body blockage scenarios according to the parameter $r_0$:
$r_0$~=~\unit[30]{cm} represents a scenario of a user operating the \ac{UE} with the \textit{hand}, e.g. operating an app, and $r_0$~=~\unit[0]{cm} represents a scenario where the \ac{UE} is held in a \textit{pocket} or as a wearable device.
These scenarios are in line with the empirical measurements made in \cite{yoo2017measurements}.
In addition, we define two random blockage scenarios according to the \ac{RB} density: \unit[0]{\acp{RB}/m$^2$} represents a scenario where there is no other blockage besides the user body (\textit{empty} scenario), and \unit[3]{\acp{RB}/m$^2$} represents a \textit{crowded} scenario (such as in a busy transportation hub or a protest march \cite{measuring2019reuters}).
The combination of these blockage scenarios give us four different scenarios: \textit{empty-hand}, \textit{empty-pocket}, \textit{crowded-hand}, and \textit{crowded-pocket}.
We set the channel parameters as measured for a car park environment according to Table \ref{tab:fading_parameters} (based on \cite[Table I]{yoo2017measurements}).
\begin{table*}[]

\centering
\caption{Channel Parameters (based on \cite[Table I]{yoo2017measurements})}
\label{tab:fading_parameters}
\begin{tabular}{l|lcrcr|crcr|cc}
\toprule
\multirow{3}{*}{Environment} & \multirow{3}{*}{UE Position} & \multicolumn{4}{|c|}{Path Loss} & \multicolumn{4}{c|}{Shadowing} & \multicolumn{2}{c}{Small-Scale Fading}\\  
 & & \multicolumn{2}{|c}{LOS} & \multicolumn{2}{c|}{NLOS} & \multicolumn{2}{c}{LOS} & \multicolumn{2}{c|}{NLOS} & \multicolumn{1}{c}{LOS} & \multicolumn{1}{c}{NLOS}\\ 
 & & \multicolumn{1}{|c}{$\nu$} & \multicolumn{1}{c}{$\ell$ (dB)} & $\nu$ & \multicolumn{1}{c|}{$\ell$ (dB)} & $\alpha$ & \multicolumn{1}{c}{$\beta$} & $\alpha$ & \multicolumn{1}{c|}{$\beta$} & \multicolumn{1}{c}{$m$} & \multicolumn{1}{c}{$m$}\\
\hline
\multirow{2}{*}{Car Park} & Hand & \multicolumn{1}{|c}{1.72} & 63.4 & \multicolumn{1}{c}{1.94} & 65.3 & \multicolumn{1}{c}{4.48} & 0.27 & \multicolumn{1}{c}{1.18} & 1.52 & \multicolumn{1}{c}{3.02} & \multicolumn{1}{c}{4.68}\\
& Pocket & \multicolumn{1}{|c}{1.70} & 59.1 & \multicolumn{1}{c}{0.61} & 88.5 & \multicolumn{1}{c}{1.96} & 0.75 & \multicolumn{1}{c}{2.80} & 0.47 & \multicolumn{1}{c}{4.21} & \multicolumn{1}{c}{2.46}\\ 
\bottomrule
\end{tabular}
\end{table*}
To evaluate the network performance, we consider the following metrics: SINR coverage and \ac{ASE}. We define the SINR coverage as the probability that the SINR at the UE is larger than the threshold $\zeta$=\unit[$5$]{dB}, i.e., $\mathsf{P}[\mathrm{SINR}>\zeta]$, as to ensure high data rate reception with low bit error rate using advanced modulation and coding schemes. The \ac{ASE} is the average spectral efficiency, $\mathsf{E}[\log(1+\mathrm{SINR})]$, divided by the cell area.

\subsection{Coverage and \ac{ASE} Profile}
\label{sec:profile}

{ In this subsection, we evaluate the effect of the inter-site distance $\delta$ (network density) on coverage and \ac{ASE} of a mmWave indoor network with ceiling-mounted \acp{AP}.
For now, we focus on the \textit{empty-hand} scenario.}
Our investigation reveals that in the ceiling-mounted \ac{AP} setup, the SINR coverage and \ac{ASE} present a non-trivial behaviour which can be classified into four regions of operation, as illustrated in 
{ Fig.~\ref{fig:cov_ase_profile-carpark-empty-hand}. 
\begin{figure}[h]
    \centering
    \includegraphics[width=\linewidth,trim={0 0cm 0 0cm},clip]{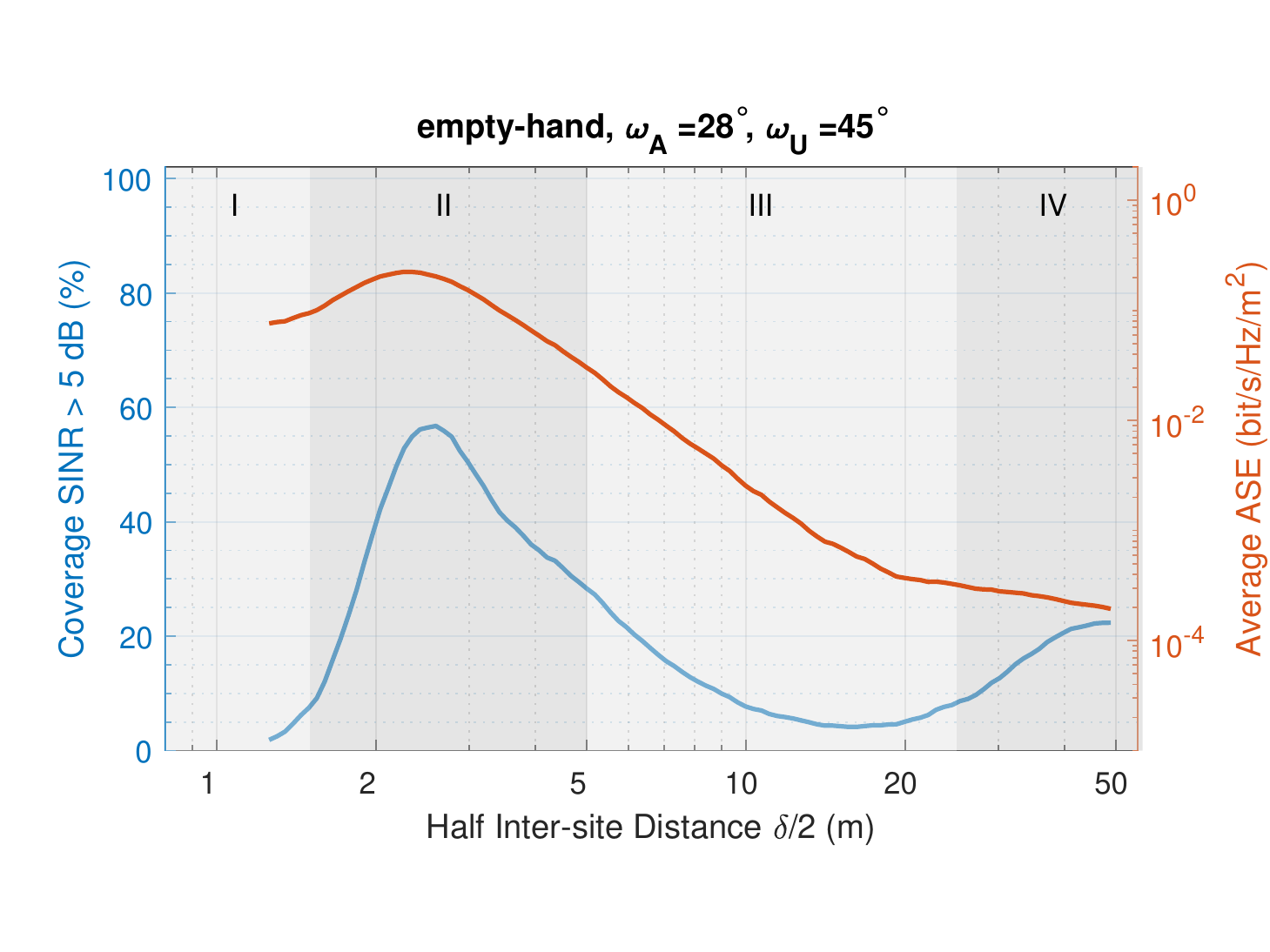}
    \caption{Coverage (left y-axis, blue curve) and \ac{ASE} (right y-axis, red curve) profile versus inter-site distance when AP beamwidth is \unit[28]{$^\circ$} and UE beamwidth is \unit[45]{$^\circ$}. The user holds the UE in hand in an empty (no other blockages) office environment and  the APs are mounted 10 m above the UE level. Regions delimited by grey rectangles represent the non-trivial behaviour of coverage with respect to AP density.}
    \label{fig:cov_ase_profile-carpark-empty-hand}
\end{figure}
These appear as we change the inter-site distance while keeping the \ac{AP} and \ac{UE} beamwidths fixed:}
\begin{tabular}{cc}
    \hfill \\
    \begin{minipage}{.6\linewidth}
    (I) \textbf{High main-lobe interference}: at high \ac{AP} density (short $\delta$), the beam is too large and causes substantial overlaps among adjacent cells, which results in high interference and, thus, low coverage. However, since the cell area is very small, the \ac{ASE} is high.
    \end{minipage} & 
    \begin{minipage}{.3\linewidth}
        \centering
        \includegraphics[width=\linewidth]{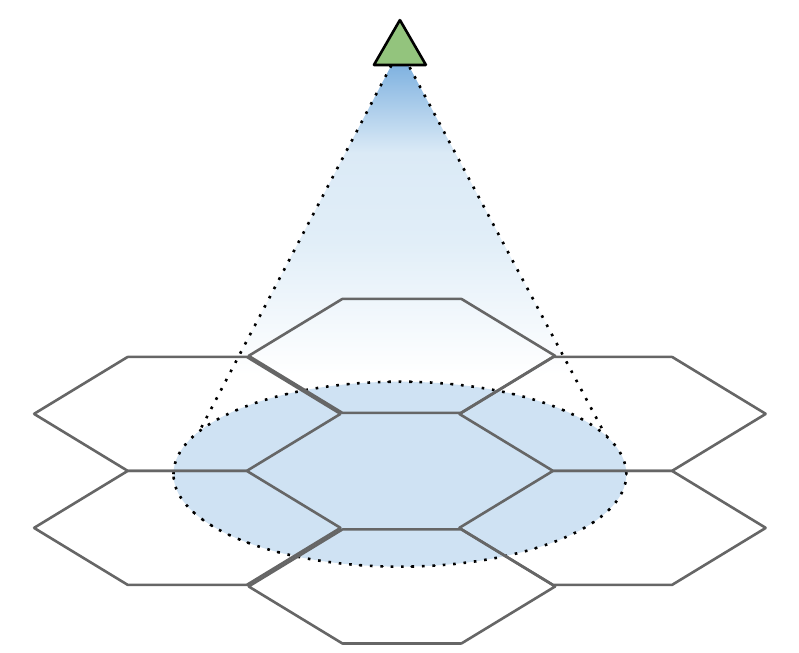}
        Region I
    \end{minipage} \\ \hfill
\end{tabular}
\begin{tabular}{cc}
    \begin{minipage}{.6\linewidth}
    (II) \textbf{Minimum main-lobe interference}: the main-lobe illuminates the entire cell with minimum interference to neighbouring cells, yielding high coverage. { However, the coverage can be limited by the \ac{UE}'s main-lobe beamwidth, as large UE beamwidths will illuminate more neighbouring \acp{AP} and increase interference power.} Thus, strong serving signal and minimised interference lead to increased SINR and to high \ac{ASE}.
    \end{minipage} & 
    \begin{minipage}{.3\linewidth}
        \centering
        \includegraphics[width=\linewidth]{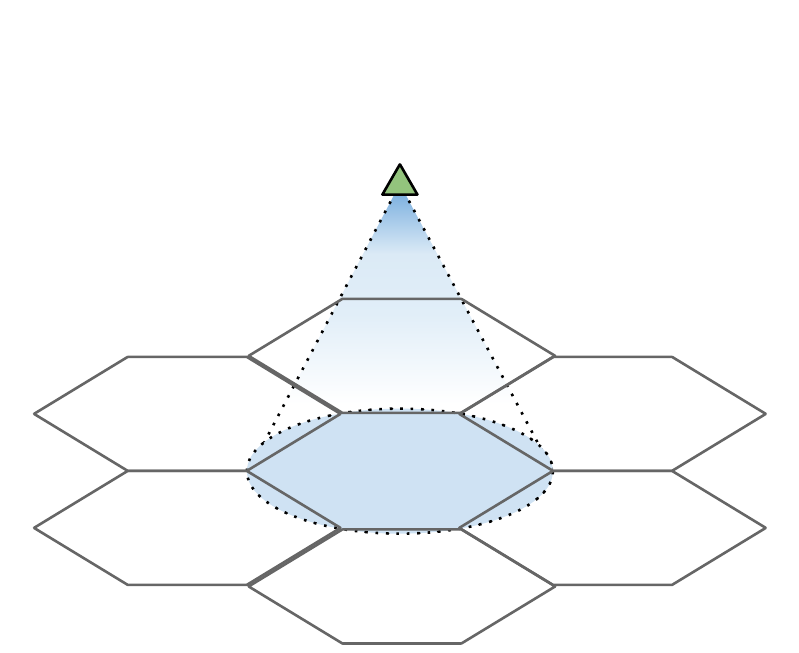}
        Region II
    \end{minipage} \\  \hfill
\end{tabular}
\begin{tabular}{cc}
    \begin{minipage}{.6\linewidth}
    (III) \textbf{Low main-lobe illumination}: at intermediate \ac{AP} densities, the coverage is very low due to the lack of main-lobe illumination by the serving \ac{AP} and due to the presence of neighbouring side-lobe interference; however, this interference decreases as the deployment gets sparser { and less \acp{AP} are illuminated by the \ac{UE} main-lobe}, leading to increased coverage. At the same time when we move towards a sparser deployment, as the cell size becomes larger, the \ac{ASE} is reduced.
    \end{minipage} & 
    \begin{minipage}{.3\linewidth}
        \centering
        \includegraphics[width=\linewidth]{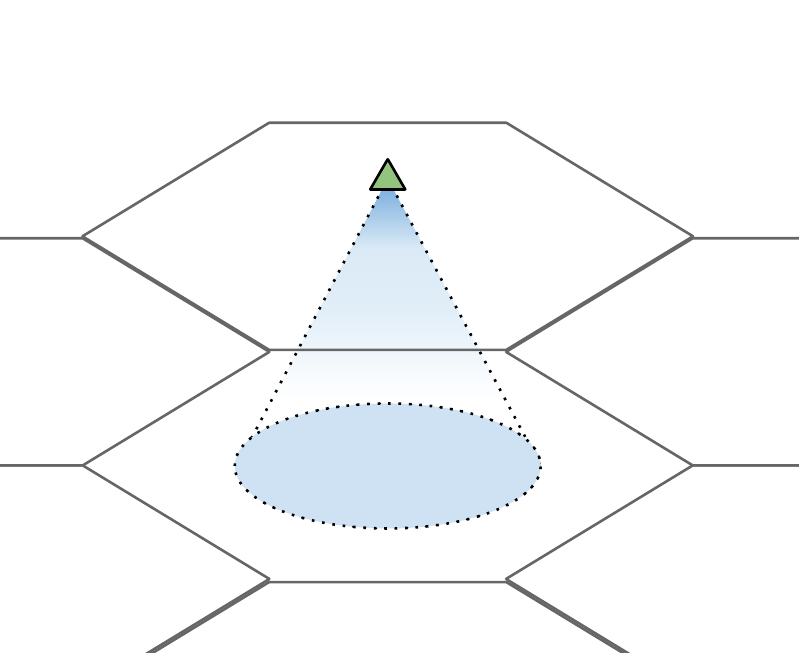}
        Region III
    \end{minipage} \\  \hfill
\end{tabular}
\begin{tabular}{cc}
    \begin{minipage}{.6\linewidth}
    (IV) \textbf{High path loss}: in low \ac{AP} density (large $\delta$), the illuminated area by the AP's beam is so small, compared to the cell area, that it becomes negligible. Therefore, the only signal that can be picked up by the majority of users comes from the side-lobes. This signal is thus weak enough for the UE receive antenna gain to play a major role in mitigating the high path loss and the AP side-lobe attenuation, and thus increasing coverage. 
    The large cell area contributes to an \ac{ASE} even lower than for the other regions.
    \end{minipage} & 
    \begin{minipage}{.3\linewidth}
        \centering
        \includegraphics[width=\linewidth]{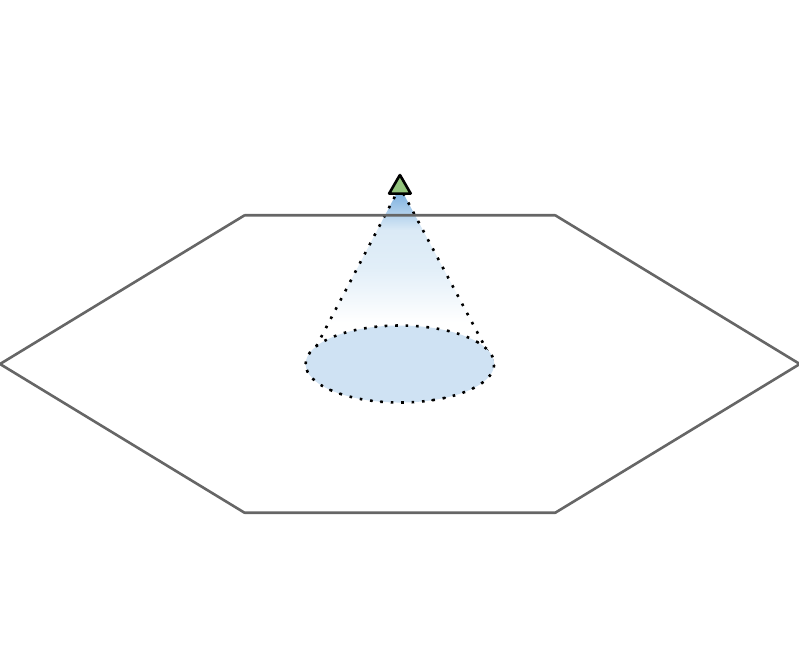}
        Region IV
    \end{minipage} \\ \hfill
\end{tabular}

\begin{figure}[h]
    \centering
    
    \includegraphics[width=\linewidth,trim={0 0cm 0 0cm},clip]{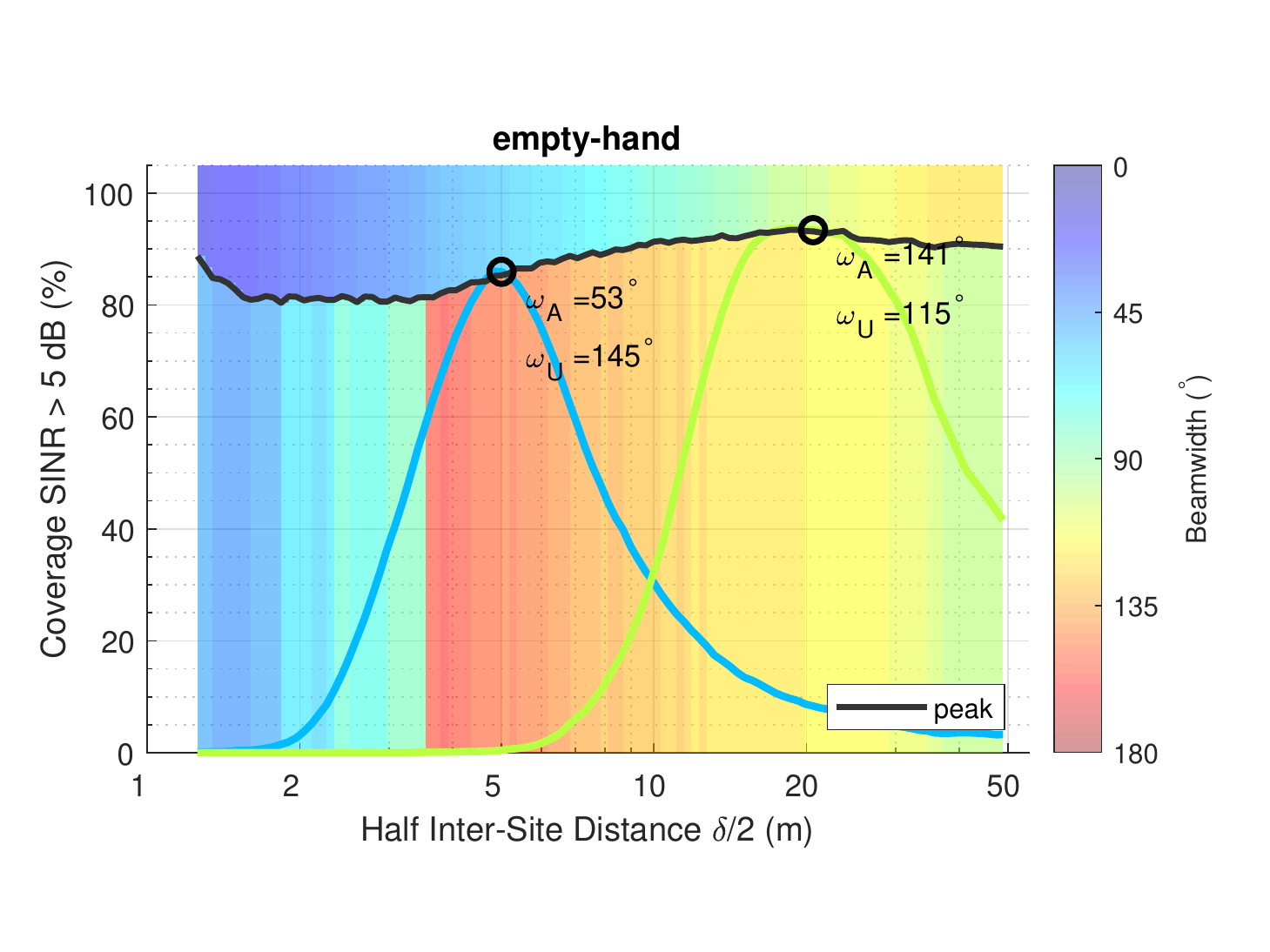}

    \caption{Peak coverage as a function of the AP density in \textit{empty-hand} scenario. The solid black line denotes the maximum performance achieved for a given inter-site distance with the AP and UE beamwidths optimised for coverage. The coloured bars above the line indicate the optimal AP beamwidth for a given inter-site distance, and the coloured bars below the line indicate the optimal UE beamwidth.}
    \label{fig:coverage2D-carpark-empty-hand}
\end{figure}

Based on these results, it is clear that a network operating in Region II reaches peak coverage performances. 
We explore the coverage behaviour for a range of AP and UE beamwidths to identify the configurations that maximise coverage for each inter-site distance $\delta$.
In Fig.~\ref{fig:coverage2D-carpark-empty-hand}, we indicate the peak coverage with a solid black line. The AP and UE beamwidth configurations that lead to that coverage, for a given $\delta$, are indicated by the colours above and below the line, respectively.
With such configurations, we observe that the peak coverage is high for any inter-site distance.

One can see the colour variation on the bars above and below the black line, where cold colours represent small beamwidths and warm colours large beamwidths.
\textit{This indicates that, for each AP density, there is a specific optimal AP and UE configuration that maximises coverage} (see Fig.~\ref{fig:beamwidths-carpark}, thin blue line, for exact numerical results).
We observe that, when increasing the cell size, larger AP beamwidths (warmer colours) are needed to achieve the peak coverage, as there is the need to illuminate the entire cell. 
However, UE beamwidth does not have the same monotonic behaviour:
With very small cell sizes ($\delta<$ \unit[2]{m}), the UE beamwidth is small (cold colours), so it does not illuminate many neighbouring APs, and thus, avoids increased interference.
As the cell size increases, the average distance between the UE and the serving AP also increases.
Hence, the UE's main-lobe tends to illuminate an increased area, as the main-lobe elevation lowers with the distance to the serving AP.
Thus, in order to avoid enhancing the signal power from many interfering APs, the UE antenna should have low directivity gain (with large beamwidths, warm colours).
Then, as the cell size becomes large ($\delta>$ \unit[5]{m}), the serving AP moves further and further away, and high directivity gain is needed to compensate for the increased path loss, decreasing the UE beamwidth.

\begin{figure}[h]
    \centering
    
    \includegraphics[width=\linewidth,trim={0 0cm 0 0cm},clip]{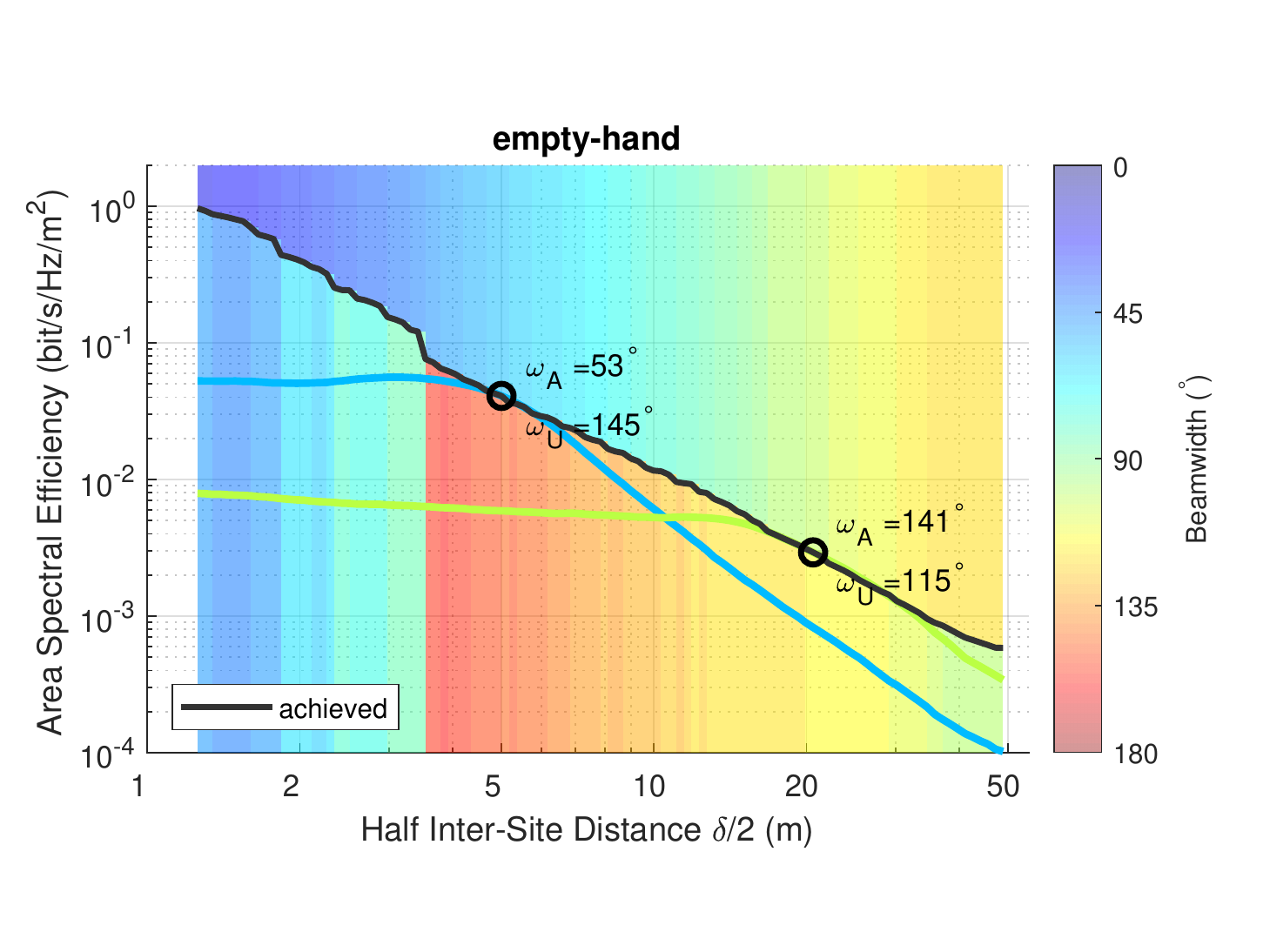}
    
    \caption{Average \ac{ASE} as a function of the AP density in \textit{empty-hand} scenario. The black solid line denotes the performance achieved for a given inter-site distance with the AP and UE beamwidths optimised for coverage. The coloured bars above the line indicate the optimal AP beamwidth for a given inter-site distance, and the coloured bars below the line indicate the optimal UE beamwidth.}
    \label{fig:ase2D-carpark-empty-hand}
\end{figure}

In Fig.~\ref{fig:ase2D-carpark-empty-hand}, we draw a solid black line that represents the \ac{ASE} achieved when using the optimal configuration for coverage.
We see that the achieved \ac{ASE} decreases due to both the increased cell area and increased path loss on the serving AP signal power.
Therefore, as peak coverage and achieved \ac{ASE} have opposite trends with respect to the cell size, there is a trade-off between these two performance metrics.
This trade-off becomes more significant when other blockage scenarios are evaluated as we shall see in the following subsection.

\subsection{Body Blockage Impact}
\label{sec:bodyblockageimpact}
\begin{figure}
    \centering
    \includegraphics[width=\linewidth]{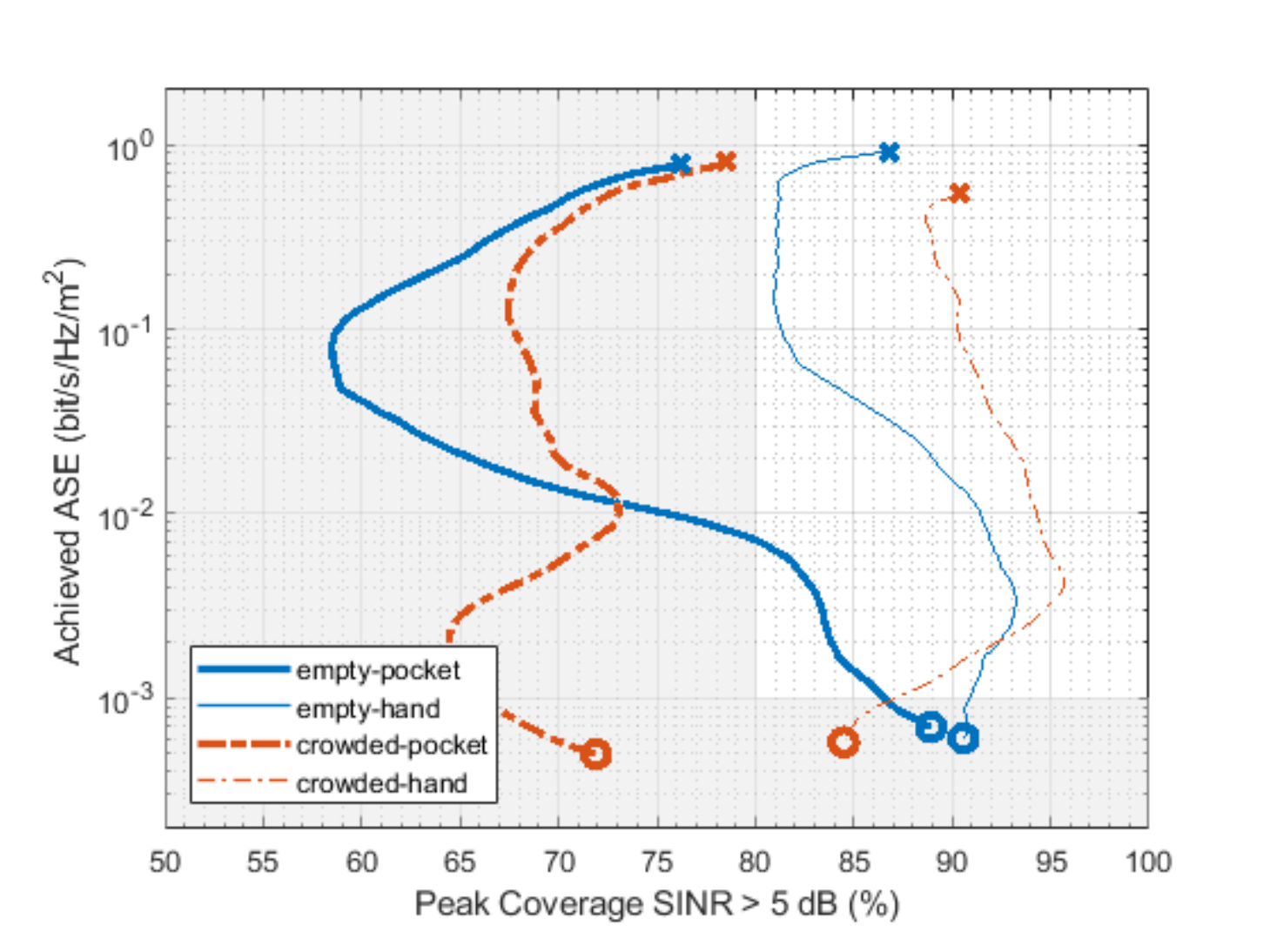}
    \caption{Body blockage impact on the ASE-coverage trade-off in the car park environment. Comparison between empty venue (blue solid lines), crowded venue (red dashed lines), UE in pocket (thick lines), and UE in hand (thin lines) scenarios.}
    \label{fig:cov_ase-carpark}
\end{figure}

In this subsection, we investigate the impact of body blockage on the ASE-coverage trade-off. To that end, we compare the four different blockage scenarios of interest: \textit{empty-hand}, \textit{crowded-hand}, \textit{empty-pocket}, and \textit{crowded-pocket}.
We depict the trade-off in Fig.~\ref{fig:cov_ase-carpark}, where each point of a curve corresponds to the optimal AP/UE beamwidths and AP density configurations that maximise coverage.
The circle \yale{$\circ$} marks the achievable ASE-coverage for the lowest AP density (largest $\delta$), while the \yale{$\times$} marks the achievable ASE-coverage for the highest AP density (smallest $\delta$). 

In all scenarios, maximum \ac{ASE} is achieved with the highest AP density, while maximum coverage can be achieved with a lower AP density.
Still, we see that coverage in \textit{hand} scenarios (thin lines) are satisfactory in highly-dense AP deployments, but in \textit{pocket} scenarios (thick lines), coverage is deeply degraded. 
For instance, if considering a coverage of \unit[80]{\%}, deploying high AP density for \textit{pocket} scenario is insufficient. To overcome this threshold, low AP density should be deployed at the expense of two orders of magnitude decrease in \ac{ASE}. 

\begin{figure}
    \centering
    \begin{subfigure}{\linewidth}
    \includegraphics[width=\linewidth,trim={0 0cm 0 0},clip]{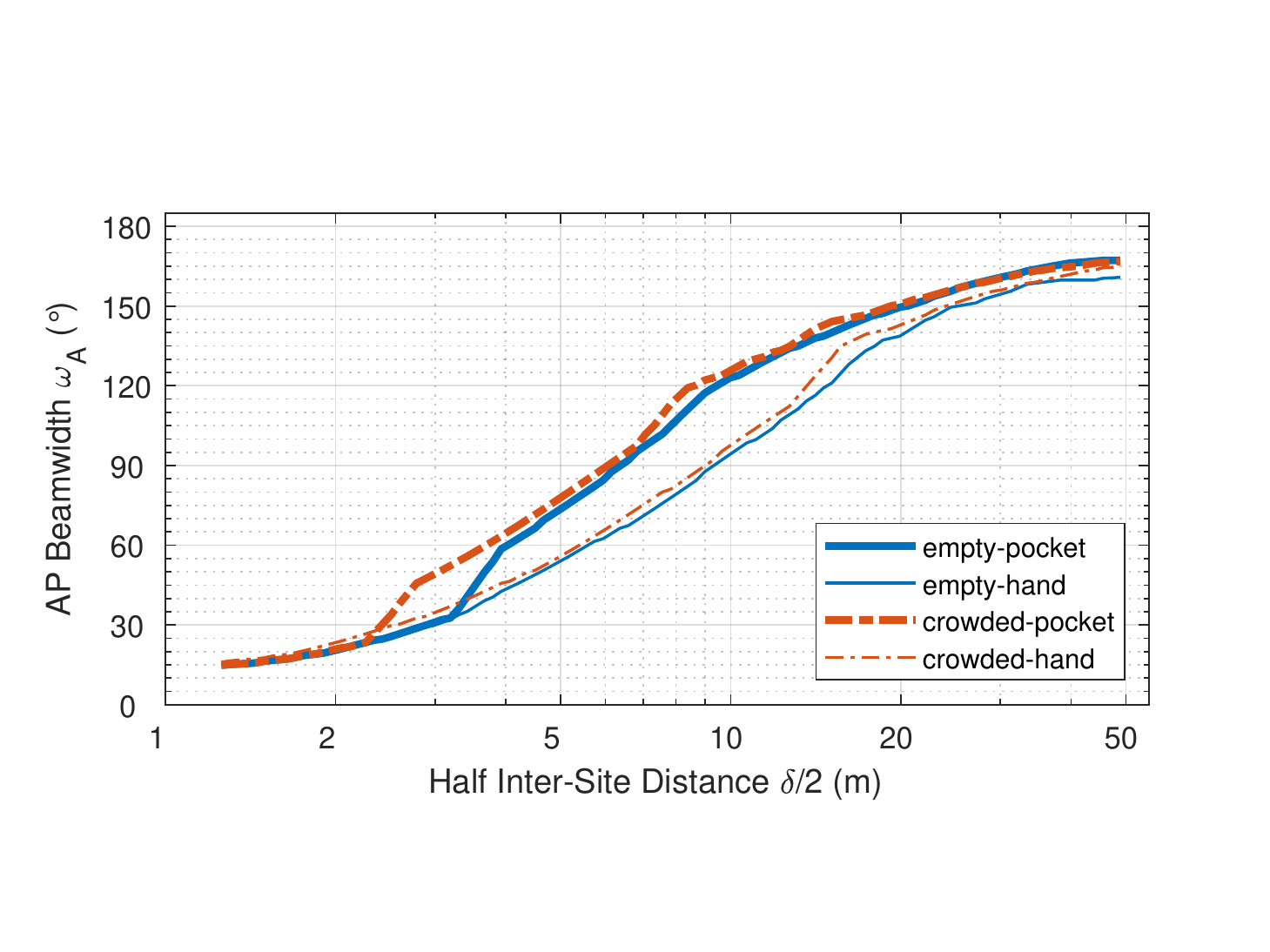}
    \caption{Optimal AP beamwidth.}
    \label{fig:ap_beamwidth-carpark}
    \end{subfigure}
    \hfill
    \begin{subfigure}{\linewidth}
    \includegraphics[width=\linewidth,trim={0 0 0 0cm},clip]{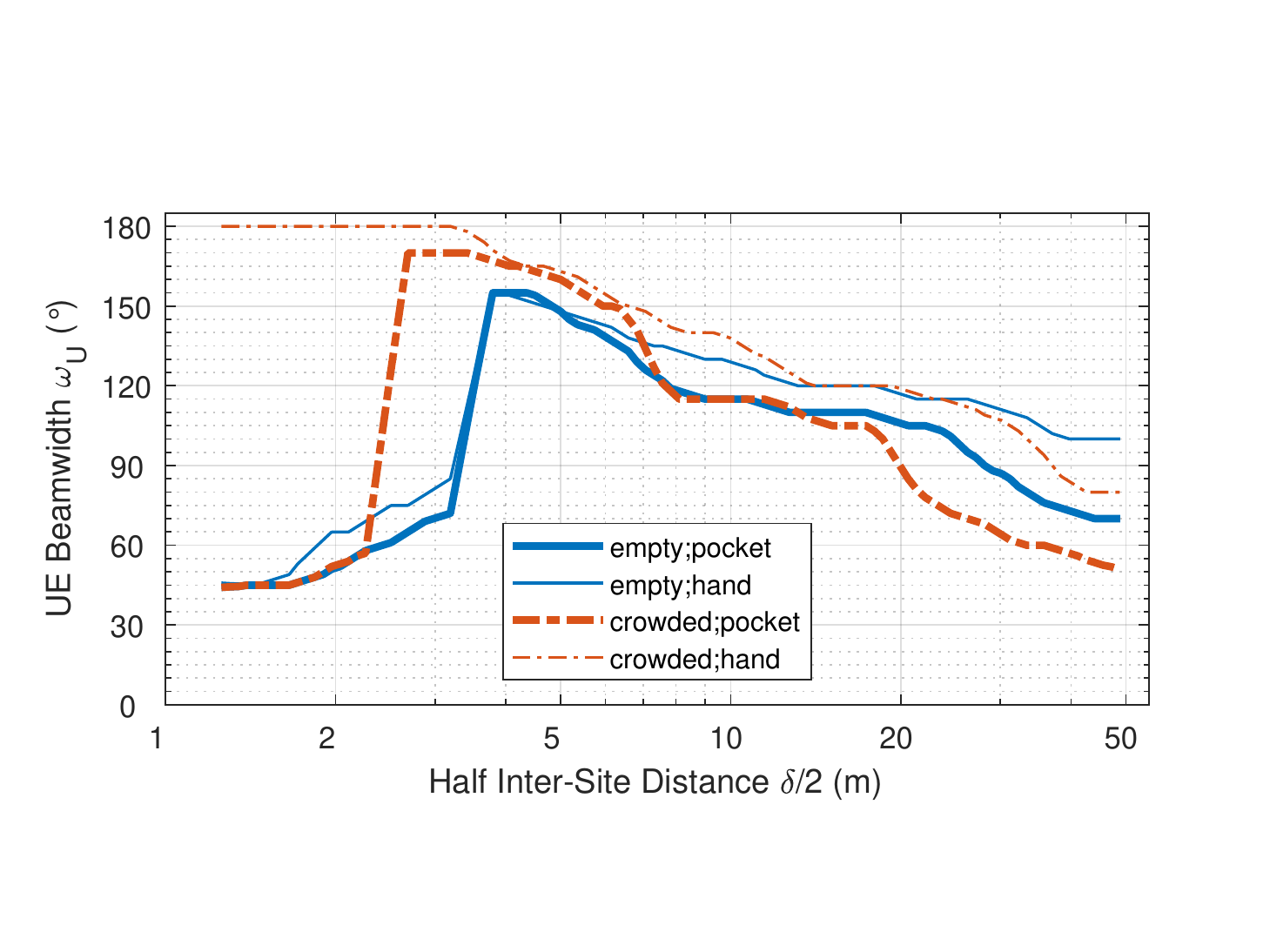}
    \caption{Optimal UE beamwidth.}
    \label{fig:ue_beamwidth-carpark}
    \end{subfigure}
    \caption{ Body blockage impact on the optimal AP and UE beamwidth in the car park environment. Comparison between empty venue (blue solid lines), crowded venue (red dashed lines), UE in pocket (thick lines), and UE in hand (thin lines) scenarios.}
    \label{fig:beamwidths-carpark}
\end{figure}

The optimal AP and UE beamwidths that lead to peak coverage for a given $\delta$ are shown in Fig.~\ref{fig:beamwidths-carpark}.
We showed in Section~\ref{sec:profile} that minimum AP's main-lobe overlap is necessary to achieve peak coverage, but now, considering a more severe body blockage, a large overlap is needed to allow the UE to be handed-off to a neighbouring AP.
\textit{It means that, for the same cell size, a larger AP beamwidth is necessary to achieve peak coverage in \textit{pocket} scenario (thick lines), than in \textit{hand} scenario (thin lines), as we observe in Fig.~\ref{fig:ap_beamwidth-carpark}.}
Yet, the larger an AP beam overlap is needed, the more interference is added, which is why the peak coverage in \textit{pocket} scenarios diminishes.
\textit{In addition, the UE beamwidth in \textit{pocket} scenarios are smaller than in \textit{hand} scenarios as the UE needs to use higher directivity gains to compensate for the increased interference, as depicted in Fig.~\ref{fig:ue_beamwidth-carpark}.}

Introducing other randomly distributed bodies, i.e., in \textit{crowded} scenarios (red lines), we observe from Fig.~\ref{fig:cov_ase-carpark} that coverage is better than in \textit{empty} scenarios when deploying high AP densities as the random bodies help to block interference.
On the other hand, with low AP densities, the random bodies create additional blockages to the serving AP, decreasing SINR.

Comparing the AP beamwidths, we do not observe significant variation between the \textit{crowded} and the \textit{empty} scenarios, as the AP beamwidth is mainly affected by the self-body blockage.
Yet, comparing the UE beamwidths, we see that the UE beamwidth in \textit{empty} scenarios (blue lines) are smaller than in \textit{crowded} scenarios (red lines), when high AP density is deployed ($\delta/2<$\unit[ 10]{m} in \textit{pocket} scenarios and $\delta/2<$\unit[ 20]{m} in \textit{hand} scenarios). This is because, differently from the \textit{crowded} scenario, there is no additional blockage to shield from the interference. Thus, the UE needs to have small beamwidths to limit a number of illuminated interfering APs.
Furthermore, we see that the optimal UE beamwidth in any scenario is not smaller than \unit[45]{$^\circ$}. \textit{It means that large beamwidths are, in general, more beneficial than very small beamwidths (i.e., \yale{pencil-beams}) in indoor ceiling-mounted AP deployments. 
This is because small UE beamwidths may still illuminate many APs and increase the interference power with high directivity gains.}

To sum up this section, the ASE-coverage trade-off becomes more significant as coverage degrades with more severe blockage scenarios. It is more difficult in a scenario with high blockage probability (\textit{pocket} scenarios) to achieve a certain level of coverage without significantly degrading the \ac{ASE}, compared to a scenario with low blockage probability (\textit{hand} scenarios).
For instance, for a requirement of \ac{ASE} \unit[$>10^{-3}$]{bit/s/Hz/m$^2$} and coverage \unit[$>80$]{\%}, the system configuration for \textit{pocket} scenarios is restricted to a range of half inter-site distances between 20 and \unit[30]{m} if the venue is \textit{empty}\footnote{ When the venue is \textit{crowded}, it is impossible to meet these design criteria.}. For \textit{hand} scenarios, the goal can be achieved with any half inter-site distance $\delta/2<$ \unit[30]{m}.
Moreover, the optimal system configuration should be set depending on the blockage scenario, as the same configuration for one scenario may not be optimal for the others.
For instance, to achieve \ac{ASE} \unit[$>10^{-1}$]{bit/s/Hz/m$^2$}, cell size and AP beamwidth should be set small (e.g., $\delta/2<$ \unit[3]{m}, $\omega_\mathrm{A}=$ \unit[30]{$^\circ$}) for all scenarios, but the optimal UE beamwidth should be large for \textit{hand} scenarios (e.g., $\omega_\mathrm{U}=$ \unit[180]{$^\circ$}) and small for \textit{pocket} scenarios (e.g., $\omega_\mathrm{U}=$ \unit[45]{$^\circ$}).

\section{Conclusion}
\label{sec:conclusions}

Herein, we studied the network design aspects of ceiling-mounted \ac{mmWave} \ac{AP} deployments with directional antennas over a confined area.
We showed that, each densification level will require individual AP and UE main-lobe beamwidths configurations to achieve maximum coverage operation.
The AP beamwidth increases with the cell size, small UE beamwidths can boost the signal from far APs in sparse deployments, while in dense deployments large UE beamwidths are more beneficial.
We also showed that the optimal AP/UE main-lobe beamwidths configuration takes different values depending on severity of the blockage scenario.
In the main, large AP beamwidths create enough overlap to allow for hand-offs that will mitigate the blockage effects while keeping interference at an acceptable level.
Finally, it is necessary for network designers to carefully consider their intended application as there is a fundamental trade-off where same AP/UE beamwidth and density configurations that optimise coverage do not optimise \ac{ASE}.


\appendices

\section{}
\label{ap:simulationcodes}
All simulation scripts used to generate the presented results were written in MATLAB\textsuperscript{\textregistered} and can be cloned from the following repository: 

\textit{https://github.com/firyaguna/matlab-nemo-mmWave}

\section{Distribution of Random Body Shadowing Angle $\Phi$}
\label{ap:distributionPhi}
\begin{wrapfigure}[13]{r}{4.5cm}
    \centering
    \includegraphics[width=.8\linewidth]{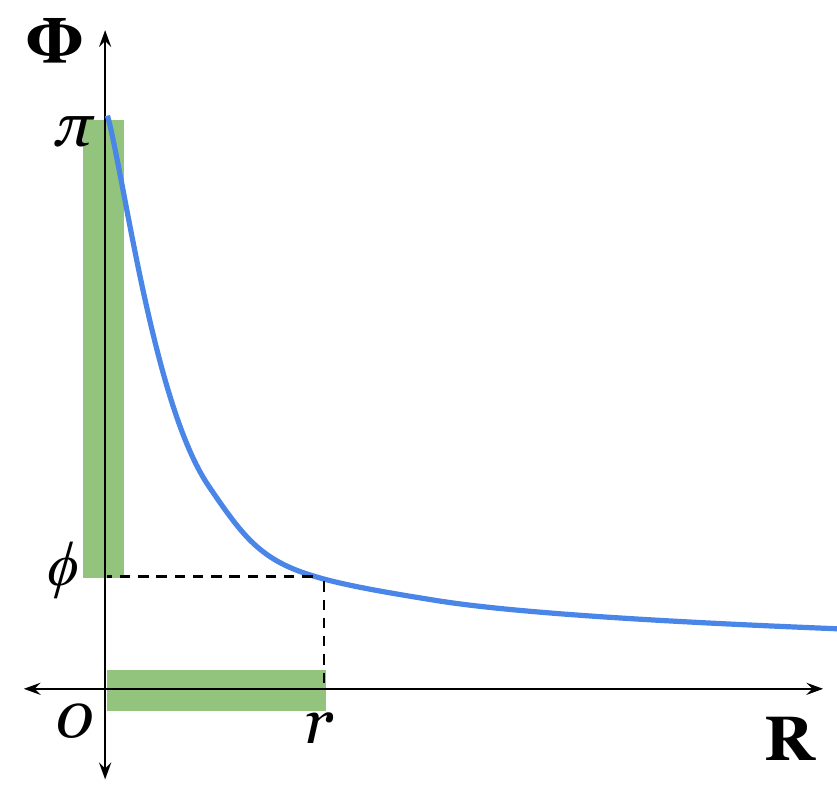}
    \caption{Body shadowing angle $\Phi$ as function $\varphi$ of the body distance $R$.}
    \label{fig:body_shadowing_angle_function}
\end{wrapfigure}
In order to compute the probability of AP being blocked by a random body in Appendix \ref{ap:proofp1}, we need to calculate the distribution of the random body shadowing angle. 
Here, since both UE and the body are uniformly positioned, the distance between them is random, which means that the body shadowing angle is also random.
Thus we define the body shadowing angle $\Phi$ as a function $\varphi(R)$ of the random distance $R$ to the body (i.e., the closer the body is to the UE, the larger the shadowing angle) and the fixed parameter --- width of the body $w_\mathrm{B}$. Then, the distribution of $\Phi$ can be obtained from the distribution of $R$ using the following change of variable:
\begin{equation}
    \Phi = \varphi(R) = 2 \arctan \frac{w_\mathrm{B}}{2R},\; R > 0. \\
    \label{eq:shadowingAngle}
\end{equation}
\begin{equation}
    R = \varphi^{-1}(\Phi) = \frac{w_\mathrm{B}}{2 \tan \frac{\Phi}{2}},\, 0 < \Phi \leq \pi. \\
    \label{eq:invShadowingAngle}
\end{equation}
\begin{equation} 
\begin{split}
    \mathsf{F}_\Phi(\phi) &= \mathsf{P}[ \Phi \leq \phi ] \\
    &= 1 - \mathsf{P}[ \phi < \Phi < \pi ] \\
    &\overset{(a)}= 1 - \mathsf{P}[ \varphi^{-1}(\phi) > \varphi^{-1}( \varphi(R) ) > \varphi^{-1}(\pi) ] \\
    &= 1 - \mathsf{P}[ \varphi^{-1}(\phi) > R > 0 ] \\
    &\overset{(b)}= 1 - \mathsf{F}_R( \varphi^{-1}(\phi) ).
\end{split} 
\label{eq:cdfPhi}
\end{equation}
where (a) follows from the change of variable in (\ref{eq:invShadowingAngle}). It should be noted that the inversion of the inequality follows from the fact that $\varphi(R)$ is a strictly decreasing function. The cdf $\mathsf{F}_R(r)$ in step (b) is the distribution of distance $R$ between two random points thrown on a square with side $s$ given by Equation (53) in \cite{moltchanov2012distance}. That equation is defined for two intervals: $r \in (0,s]$ and $r \in (s,s\sqrt{2})$. Here, we consider only the first interval $r \in (0,s]$, as we assume that blockages from very far bodies (with distance $r \in (s,s\sqrt{2})$) are negligible.
Then, the pdf is the derivative of the cdf in (\ref{eq:cdfPhi}):
\begin{equation}
\begin{split}
    \mathsf{f}_\Phi(\phi) &= \frac{\mathrm{d}\mathsf{F}_\Phi(\phi)}{\mathrm{d}\phi} = -\frac{\mathrm{d}\mathsf{F}_R(\varphi^{-1}(\phi))}{\mathrm{d}\phi} \\
    &\overset{(a)}= -\frac{\mathrm{d}\mathsf{F}_R(r)}{\mathrm{d}\phi} = -\frac{\mathrm{d}\mathsf{F}_R(r)}{\mathrm{d}r} \frac{\mathrm{d}r}{\mathrm{d}\phi} \\
    &= -\left( \frac{2 \pi r}{s^2} - \frac{8 r^2}{s^3} + \frac{2 r^3}{s^4} \right) \frac{w_\mathrm{B}}{2(\cos\phi-1)} \\
    &\overset{(b)}= -\left( \frac{\pi \rho}{s^2} - \frac{4 \rho^2}{s^3} + \frac{\rho^3}{s^4} \right) \frac{w_\mathrm{B}}{(\cos\phi-1)},
    \label{eq:pdf_phi}
\end{split}
\end{equation}
where (a) follows from the change of variable in (\ref{eq:invShadowingAngle}), and (b) follows from the change of variable in (\ref{eq:shadowingAngle}), with $\rho$ as a short notation of the function $\varphi^{-1}(\phi) = w_\mathrm{B}/(2\tan\frac{\phi}{2})$.

\section{Probability of Self-body Blockage}
\label{ap:proofp0}
The event [\textit{the AP is blocked by user body}] is a specific case of the event given by Definition \ref{def:bodyblockage}, where we know that the user body is at a fixed distance $r_0$ from the UE and its orientation is a random variable $\Theta$ uniformly distributed in $[0,2\pi)$.
Then, the event [\textit{the AP is blocked by user body}] is given by:
\begin{equation}
    [\Theta < \phi_0] \cap [d_\mathrm{A} \geq z_0],
\end{equation}
where $z_0 = r_0 \cdot \frac{h_\mathrm{A}}{h_\mathrm{B}}$ is self-body blockage free zone and $\phi_0=\varphi(r_0)$ is the self-body shadowing angle.
Therefore, since the body orientation $\Theta$ is independent of the distance $r_0$, the probability of self-body blockage is:
\begin{equation}
\begin{split}
    p_0 &= \mathsf{P}[\Theta \leq \varphi(r_0) \; \cap \; r_0 \leq d_\mathrm{A}\cdot {h_\mathrm{B}}/{h_\mathrm{A}} ]\\
        &= \mathsf{P}[\Theta \leq \varphi(r_0)] \; \mathbbm{1}_{(r_0 \leq d_\mathrm{A}\; {h_\mathrm{B}}/{h_\mathrm{A}}]}\\
\end{split}
\end{equation}
It means that if the UE is at a distance { $d_\mathrm{A} \leq r_0\cdot {h_\mathrm{A}}/{h_\mathrm{B}}$}, the probability of self-body blockage is zero (UE inside the self-blocking free zone). Otherwise, the probability is $\mathsf{F}_\Theta(\varphi(r_0))$. Thus, we can express $p_0$ as:
\begin{equation}
\begin{split}
    p_0(d_\mathrm{A}) &= \mathsf{F}_\Theta(\varphi(r_0)) \cdot \mathbbm{1}_{(r_0 \leq d_\mathrm{A}\; {h_\mathrm{B}}/{h_\mathrm{A}}]} \\
        &= \frac{\varphi(r_0)}{2\pi} \cdot \mathbbm{1}_{(r_0 \leq d_\mathrm{A}\; {h_\mathrm{B}}/{h_\mathrm{A}}]} \\
        &= \left\{
    \begin{split}
        & \frac{\arctan(w_\mathrm{B}/_{2r_0})}{\pi}, \; d_\mathrm{A} > r_0 \cdot {h_\mathrm{A}}/{h_\mathrm{B}}; \\
        & 0,\;\; 0 < d_\mathrm{A} \leq r_0 \cdot {h_\mathrm{A}}/{h_\mathrm{B}}.
    \end{split}
    \; \right.
\end{split}
\end{equation}

\section{Probability of AP Being Blocked by a Random Body}
\label{ap:proofp1}
Assuming both \ac{UE} and an \ac{RB} (not the user) positions are uniformly distributed, the event [\textit{the body is in between the AP and UE}] depends on the distance $R$ from the RB to the UE, on the body shadowing angle $\Phi$ (these random variables are discussed in Appendix \ref{ap:distributionPhi}), and on the body orientation $\Theta$, which we also assume to be uniformly distributed in $[0,2\pi)$ for an \ac{RB}.
Then, applying the Definition \ref{def:bodyblockage} to a random body, the event [\textit{the AP is blocked by a random body}] is given by:
\begin{equation}
    [\Theta < \Phi] \cap [d_\mathrm{A} \geq Z],
\end{equation}
where $Z = R \cdot \frac{h_\mathrm{A}}{h_\mathrm{B}}$ is the  blockage free zone and $\Phi=\varphi(R)$ is the body shadowing angle.
Therefore, the probability of blockage by a random body is:
\begin{equation} 
\begin{split}
    p_1(d_\mathrm{A}) &= \mathsf{P}[ \Phi > \Theta \; \cap \; 0 < R \leq d_\mathrm{A}\cdot {h_\mathrm{B}}/{h_\mathrm{A}} ] \\
    &\overset{(a)}= \mathsf{P}[ \Phi > \Theta \; \cap \; \varphi( d_\mathrm{A}\cdot {h_\mathrm{B}}/{h_\mathrm{A}} ) < \Phi < \varphi(0) ] \\
    &= \mathsf{P}[ \Phi > \Theta \; \cap \; \varphi( d_\mathrm{A}\cdot {h_\mathrm{B}}/{h_\mathrm{A}} ) < \Phi < \pi ] \\
    &= \int\limits_{\varphi(d_\mathrm{A}\cdot {h_\mathrm{B}}/{h_\mathrm{A}})}^\pi \int\limits_{0}^\phi \mathsf{f}_{\Theta,\Phi}(\theta,\phi) \; \mathrm{d}\theta\mathrm{d}\phi \\
    &\overset{(b)}= \int\limits_{\varphi(d_\mathrm{A}\cdot {h_\mathrm{B}}/{h_\mathrm{A}})}^\pi \int\limits_{0}^\phi \mathsf{f}_{\Theta}(\theta) \mathrm{d}\theta \; \mathsf{f}_{\Phi}(\phi) \mathrm{d}\phi \\
    &= \int\limits_{\varphi( {d_\mathrm{A}}\cdot {h_\mathrm{B}}/{h_\mathrm{A}} )}^\pi \mathsf{F}_\Theta(\phi) \mathsf{f}_\Phi(\phi) \; \mathrm{d}\phi\\
    &\overset{(c)}= \int\limits_{\varphi( {d_\mathrm{A}}\cdot {h_\mathrm{B}}/{h_\mathrm{A}} )}^\pi \frac{\phi}{2\pi} \frac{-w_\mathrm{B}}{(\cos\phi-1)} \left( \frac{\pi \rho}{s^2} - \frac{4 \rho^2}{s^3} + \frac{\rho^3}{s^4}  \right) \; \mathrm{d}\phi,
\end{split} 
\end{equation}
where (a) follows from making a change of variables, (b) follows from the independence of $\Theta$ and $R$, and consequently, of $\Theta$ and $\Phi$, and (c) follows from substituting the cdf of $\Theta$ and the pdf of $\Phi$ with (\ref{eq:pdf_phi}) given in Appendix \ref{ap:distributionPhi}.




\bibliographystyle{IEEEtran}
\bibliography{main}

%



%

\begin{IEEEbiography}[{\includegraphics[width=1in,height=1.25in,clip,keepaspectratio]{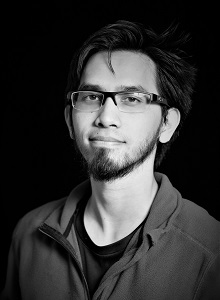}}]{Fadhil Firyaguna}
is a PhD candidate researcher at CONNECT, the Science Foundation Ireland Research Centre for Future Networks, at Trinity College Dublin.
He received his MSc degree in Electronics and Automation Systems Engineering from the University of Brasilia in 2014 and a BSc degree in Communication Networks Engineering in 2012, also from the University of Brasilia.
He has experience in telecommunication systems and his research interests focus on 5G cellular networks, millimetre-wave communications, ad hoc networks, discrete-event simulations, medium-access control (MAC) protocols, and MIMO communications.
\end{IEEEbiography}
\begin{IEEEbiography}[{\includegraphics[width=1in,height=1.25in,clip,keepaspectratio]{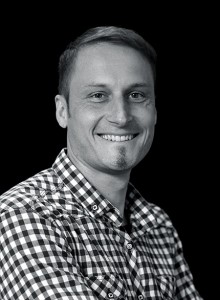}}]{Jacek Kibi{\l}da}
is a Research Fellow with Trinity College Dublin, Ireland, and a Challenge Fellow with Science Foundation Ireland. He received his PhD degree from Trinity College Dublin in 2016, and in 2017-18 he has been the Fulbright Scholar. His research interests include applications of stochastic geometry and optimization to modelling of radio resource management problems in wireless and mobile networks.
\end{IEEEbiography}

\begin{IEEEbiography}[{\includegraphics[width=1in,height=1.25in,clip,keepaspectratio]{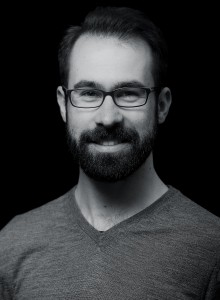}}]{Carlo Galiotto}
received his PhD in Electrical and Electronic Engineering from Trinity College Dublin, Ireland, in 2017.
He received his M.Sc. in Telecommunications Engineering from University of Padova, Italy, in 2009. After that, he spent two years as a researcher at Aalborg University, Denmark, where he worked on passive RFID systems and on radio resource management for heterogeneous cellular networks. 
His work focuses on the study of performance and issues of extremely dense networks, design and optimization of urban small-cell networks, as well as spectrum sharing for 5G systems.
\end{IEEEbiography}

\begin{IEEEbiography}[{\includegraphics[width=1in,height=1.25in,clip,keepaspectratio]{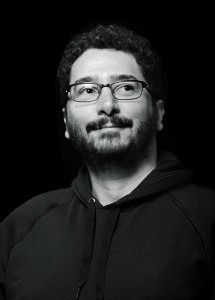}}]{Nicola Marchetti}
is currently Assistant Professor in Wireless Communications at Trinity College Dublin, Ireland. He performs his research under the Irish Research Centre for Future Networks and Communications (CONNECT), where he leads the Wireless Engineering and Complexity Science (WhyCOM) lab.
He received the PhD in Wireless Communications from Aalborg University, Denmark in 2007, and the M.Sc. in Electronic Engineering from University of Ferrara, Italy in 2003. He also holds an M.Sc. in Mathematics which he received from Aalborg University in 2010. His research interests include Radio Resource Management, Self-Organising Networks, Complex Systems Science and Signal Processing for communication networks. He has authored in excess of 130 journals and conference papers, 2 books and 8 book chapters, holds 4 patents, and received 4 best paper awards. He is a senior member of IEEE and serves as an associate editor for the IEEE Internet of Things Journal since 2018 and the EURASIP Journal on Wireless Communications and Networking since 2017.
\end{IEEEbiography}




\end{document}